\documentclass[11pt]{article}
\usepackage[margin=1in]{geometry}
\usepackage{microtype}
\usepackage{setspace}
\usepackage{amsmath,amssymb,amsthm}
\usepackage{booktabs,longtable,array,threeparttable,calc}
\usepackage{graphicx}
\graphicspath{{figures/}}
\usepackage{caption}
\usepackage{float}
\usepackage{placeins}
\usepackage{enumitem}
\usepackage{natbib}
\usepackage[hidelinks]{hyperref}
\usepackage[nameinlink,noabbrev]{cleveref}
\usepackage{pdflscape}
\usepackage{xcolor}
\usepackage{siunitx}
\usepackage{ragged2e}
\newtheorem{proposition}{Proposition}
\title{\textbf{Practice Makes Perfect:}\\[0.4em]\large Tacit Knowledge, Generative AI, and Returns to Schooling and Experience}
\author{Eduardo Levy Yeyati\thanks{Universidad Torcuato Di Tella and IDB.} \and Dar\'io Judzik\thanks{Universidad Torcuato Di Tella.} \and Juan Ignacio Barbiero\thanks{Universidad Torcuato Di Tella.}}
\date{First draft: September 2026 \\ This version: October 2026}
\begin{document}
\maketitle
\begin{abstract}
Generative AI changes the scarcity of different forms of human capital. It makes codified knowledge cheaply available at the point of production, while judgment acquired through practice remains tied more closely to the worker. We ask what this does to two familiar prices in the wage structure: the return to schooling and the return to experience. In a short-run framework, machine-supplied codified knowledge substitutes directly for the codified component of worker human capital, lowering the schooling gradient. If codified and practice-based knowledge are complements, the return to experience rises where learning through practice is fastest; technology-specific experience can also become obsolete, rotating the experience profile rather than shifting it uniformly. Using US Current Population Survey earnings data from 2018 through June 2026, we compare occupations in the top and bottom quartiles of the pre-ChatGPT Felten--Raj--Seamans exposure index. After the public release of ChatGPT, the return to a year of schooling falls by 0.55 log points in exposed occupations. The return to a year of experience rises by 0.26 log points at three years of potential experience, is unchanged at fifteen, and falls by 0.12 log points at twenty-five; under the baseline quadratic, the effect crosses zero at 18.1 years. These breaks appear with the release, disappear under a pandemic placebo, survive a common censoring rule across the full sample, and remain when each occupation is allowed its own pre-shock schooling and experience gradients. The pattern is a repricing of human capital inside exposed occupations: codifiable knowledge becomes less scarce, while the value of experience depends increasingly on whether it transfers to the new technology.
\end{abstract}
\noindent\textbf{JEL codes:} J24, J31, O33, I26\\
\textbf{Keywords:} tacit knowledge; human capital; generative artificial intelligence; return to experience; return to schooling; skill prices; wage structure

\clearpage
\section{Introduction}

Generative AI does not simply change how much labour firms need. It can change what kind of human capital is scarce. A worker's expertise combines knowledge that can be codified and transmitted with judgment that is acquired through practice and remains difficult to separate from the worker who learned it. A technology that suddenly makes the first component abundant should therefore change the wage structure even if average wages move little. This paper asks whether that repricing is already visible in the returns to schooling and experience.

The distinction matters because schooling and experience load differently on these two forms of knowledge. Schooling is an imperfect proxy for explicit, codifiable knowledge: material that can be written down, taught, stored, and increasingly supplied by generative systems. Experience is an imperfect proxy for practice-based knowledge: contextual judgment, routines, and pattern recognition accumulated through repeated use. In Polanyi's (1966) terms, part of what an experienced worker knows is tacit. A caseworker learns which clients require closer follow-up; a surgeon learns when tissue tension is wrong; a negotiator learns how to read a room. Such knowledge is not reducible to a stored answer even when codified information relevant to the task becomes cheap.

The two margins need not move together. Machine-supplied codified knowledge competes directly with the codified component of worker human capital, lowering its scarcity value and the wage gradient associated with schooling. At the same time, making codified knowledge more abundant can raise the value of judgment that complements it. Since learning through practice is fastest early in a career, that force is strongest at low levels of experience. Some experience, however, is tied to the task technology on which it was acquired. When those routines lose value, the wage-experience profile can become steeper early in the career and flatter later on.

We formalise these ideas in a short-run framework in which the codified knowledge a worker acquires through schooling and the codified knowledge supplied by machines are pooled into a single composite input, while practice-based knowledge acquired through experience enters separately. The schooling result follows from substitution within that composite: as AI expands the machine-supplied component, the worker's share of codified knowledge falls, and with it the return to schooling, even when the outer production technology is Cobb--Douglas. The experience result requires an additional condition. If codified and practice-based knowledge are gross complements, expanding the former raises the marginal value of the latter. Allowing part of experience to be specific to the pre-AI task technology can, under the conditions set out in Section~\ref{sec:theory}, yield a unique crossover point in the career at which the gains from complementarity give way to obsolescence.

We test these predictions using the US Current Population Survey Outgoing Rotation Group from 2018 through June 2026. Treatment is assigned at the occupation level using the AI Occupational Exposure index of Felten, Raj and Seamans (2021), constructed before the public release of ChatGPT. We compare workers in the top and bottom quartiles of the exposure distribution before and after November 2022. The estimating equation interacts schooling and a quadratic in potential experience with treatment and post-release indicators. The parameters of interest are therefore differences-in-differences in \emph{skill prices}, not in wage levels. Identification requires parallel trends in the schooling and experience gradients; occupation and month effects, together with a treated-by-post term, absorb permanent occupation differences, common time shocks, and differential level shifts in wages.

The two prices move differently after the release. The return to a year of schooling falls by 0.55 log points in exposed occupations relative to the change in controls. Against a pre-shock return of 7.5 log points in those occupations, this represents a decline of about 7.3 percent. The experience profile, by contrast, rotates. The return to a year of experience rises by 0.26 log points at three years of potential experience, shows no statistically significant change at fifteen, and falls by 0.12 log points at twenty-five. Under the quadratic specification, the effect on the experience gradient changes sign at 18.1 years; this is not a crossover in wage gains and losses. The two ends of the profile move in opposite directions in every post-release year, as one would expect if transferable practice becomes more valuable early in the career while technology-specific experience depreciates later on.

The timing is also informative. The event studies are flat before the release for schooling and early-career experience, and the schooling break appears in the first post-release year. A placebo dated at the onset of the pandemic produces no effect; dropping the technology sector does little; and the estimated total wage loss increases with exposure. Two harder checks leave the headline results essentially unchanged. Imposing the same artificial censoring rule throughout 2018--2026 yields a schooling effect of -0.53 log points and nearly the same experience rotation, which reduces concerns that the result is driven mechanically by the change in Census top-coding rules. Allowing every occupation its own pre-shock schooling, experience, and experience-squared gradients gives -0.58, +0.26, and -0.11 log points at the same three margins. The results are therefore not inherited from permanent cross-occupation differences in Mincer slopes.

Our contribution lies mainly in the object we estimate. The human-capital literature has long studied the level and life-cycle evolution of the schooling and experience premia, while the recent AI literature has focused on productivity, employment, earnings, and knowledge formation. Here the Mincer slopes themselves are outcomes: we ask how the two skill prices move around a dated technology shock that changes the relative abundance of codified knowledge. The joint movement matters. A lower schooling gradient, a steeper early-career experience gradient, and a flatter late-career one are consistent with three distinct mechanisms in the framework: direct substitution between human and machine codified knowledge, complementarity with practice-based judgment, and obsolescence of technology-specific human capital.

Section~\ref{sec:theory} develops the framework. Section~\ref{sec:literature} places the argument in the human-capital, task, and AI literatures; Section~\ref{sec:data} describes the data and empirical design; and Section~\ref{sec:results} reports the estimates and identification checks. Section~\ref{sec:conclusion} concludes.

\section{Theoretical framework}\label{sec:theory}

The framework is deliberately short-run. Schooling and accumulated
experience are treated as predetermined over the post-2022 window, so
the objects of interest are their market prices: the percentage wage
gain associated with an additional year of schooling and the slope of
the wage-experience profile. Workers' subsequent investment choices and
the intergenerational formation of skills are left outside the model.

The central distinction is between codified knowledge and practice-based
knowledge. Schooling is an imperfect proxy for the first, while
potential experience is an imperfect proxy for exposure to the second.
Generative AI expands the effective supply of codified knowledge. When
codified and practice-based knowledge are complements, this reduces the
share of wages paid to the former and raises the share paid to the
latter. Experience acquired with older technologies may also lose value.
Together, these two forces produce a decline in the return to schooling
and a rotation of the experience profile in favour of early-career
workers.

\subsection{Human capital and production}

Worker $i$ is employed in occupation $o$ at date $t$. Let $S_i$ denote years of
schooling and $X_i$ potential labour-market experience. Schooling produces
a stock of explicit, codifiable knowledge
\begin{equation}
E_i=\phi(S_i), \qquad \phi'(S_i)>0,
\label{eq:E}
\end{equation}
while experience provides repeated opportunities to accumulate tacit
knowledge through practice, feedback, and contextual judgment. We write
this transferable practice-based stock as
\begin{equation}
T_i=\tau(X_i), \qquad \tau'(X_i)>0.
\label{eq:T}
\end{equation}

Potential experience is not itself tacit knowledge. Equation~\eqref{eq:T} is a
measurement bridge. Within an occupation, more time since leaving school
is assumed to provide, on average, greater exposure to practice. The
mapping may be noisy because time out of the labour force and
differences in task assignments are not observed. The empirical
coefficients should therefore be interpreted as returns to potential
experience, which the framework uses as a proxy for practice-based human
capital.

Let $A_{ot}$ denote the capability of generative AI available for codifiable
tasks in occupation $o$ at date $t$, and let $M(A_{ot})$ be the stock of
machine-supplied codified knowledge it makes available, with $M'(A)>0$.
Human and machine codified knowledge are perfect substitutes inside the
composite codified input $Z_{iot}=E_i+M(A_{ot})$. Output is
\begin{equation}
Y_{iot}=Q_{ot}F(Z_{iot},T_i), \qquad Z_{iot}=E_i+M(A_{ot}),
\label{eq:production}
\end{equation}
where $Q_{ot}$ collects demand and productivity factors that shift wage
levels. In the empirical implementation, these level effects are
represented by occupation effects, calendar-time effects, and a
differential treated-group shift after the release. We assume that a
worker's pay is proportional to the value of that worker's contribution,
up to a factor common to workers in the same occupation and period. That
factor can reflect output prices, firm-level diminishing returns, or a
common wage markdown; what matters for our results is that it does not
vary systematically with schooling or experience. Consequently,
\begin{equation}
\ln w_{iot}=a_{ot}+\ln F\!\left(\phi(S_i)+M(A_{ot}),\tau(X_i)\right),
\label{eq:logwage}
\end{equation}
where $a_{ot}$ absorbs $\ln Q_{ot}$ and any such common wedge between a
worker's contribution and pay. We interpret $F$ as net of AI-service
costs that do not vary systematically with the worker's schooling or
experience. The term $a_{ot}$ makes clear why the theory concerns skill
prices rather than average wages. Generative AI may raise or lower the
level of wages in an exposed occupation, and the empirical specification
allows for this through the treated-by-post term. The predictions below
concern how the slope of log wages with respect to schooling and
experience changes.

For transparency, take $F$ to be a constant-returns CES technology,
\begin{equation}
F(Z,T)=\left[\alpha Z^{\rho}+(1-\alpha)T^{\rho}\right]^{1/\rho}, \qquad Z=E+M(A),
\label{eq:ces}
\end{equation}
where $0<\alpha<1$, $\rho=(\sigma-1)/\sigma$, and $\sigma$ is the
elasticity of substitution between the codified composite $Z$ and
practice-based knowledge $T$. We focus on $\sigma<1$, equivalently $\rho<0$,
so that codified knowledge and practice-based knowledge
are gross complements. This is a maintained condition for the experience
prediction rather than a property of the exposure indices. The schooling
prediction is stronger: because AI supplies codified knowledge that is
directly substitutable for worker-supplied codified knowledge inside $Z$,
a decline in the schooling gradient can arise even when the outer
technology is Cobb--Douglas.

Define the output elasticities of the two composite inputs as
\begin{equation}
\theta_Z\equiv\frac{ZF_Z}{F}=\frac{\alpha Z^{\rho}}{\alpha Z^{\rho}+(1-\alpha)T^{\rho}},
\qquad
\theta_T\equiv\frac{TF_T}{F}=1-\theta_Z.
\label{eq:shares}
\end{equation}
Let $\eta_A\equiv\partial\ln Z/\partial\ln A>0$ denote the elasticity of the
codified composite with respect to AI capability. Differentiating the
CES shares gives
\begin{equation}
\frac{\partial\theta_Z}{\partial\ln A}=\rho\theta_Z\theta_T\eta_A<0,
\qquad
\frac{\partial\theta_T}{\partial\ln A}=-\rho\theta_Z\theta_T\eta_A>0.
\label{eq:sharechanges}
\end{equation}

When the two composite inputs are complements, a larger supply of
codified knowledge raises the output share of practice-based
knowledge. Separately, within the codified composite, the machine-supplied
component reduces the scarcity of worker-supplied codified
knowledge. The two margins generate distinct predictions for the
schooling and experience gradients.

\subsection{The return to schooling}\label{sec:schooling-theory}

The return to an additional year of schooling in the empirical wage
equation is the schooling semi-elasticity. Let $h\equiv E/Z$ denote the human
share of the codified composite. Then
\begin{equation}
r_S(A,S,X)\equiv\frac{\partial\ln w}{\partial S}
=\theta_Z(A,S,X)h(A,S)\frac{\phi'(S)}{\phi(S)}.
\label{eq:rs}
\end{equation}

Holding the worker's schooling and experience stocks
fixed over the short adjustment window, the machine contribution expands
$Z$ and reduces $h$. Using $\partial\ln h/\partial\ln A=-\eta_A$,
\begin{equation}
\frac{\partial r_S}{\partial\ln A}=r_S\eta_A(\rho\theta_T-1)<0.
\label{eq:schoolingchange}
\end{equation}

\begin{proposition}[Declining schooling return]\label{prop:schooling}
If AI supplies codified knowledge that is directly substitutable for worker-supplied codified knowledge inside $Z$, then an increase in AI capability lowers the log-wage return to an additional year of schooling for any admissible CES elasticity of substitution.
\end{proposition}

The result does not say that educated workers become less productive. AI
can raise their wage level while compressing the wage difference
associated with an additional year of schooling. There are now two
channels. First, machine-supplied codified knowledge directly dilutes
the human share $h$ of the codified input. Second, under gross
complementarity, the expansion of $Z$ shifts the outer production share
toward $T$. If $\phi(S)=\exp(\lambda S)$, equation~\eqref{eq:rs} becomes
\begin{equation}
r_S=\lambda\theta_Z h.
\label{eq:schoolingloglinear}
\end{equation}

This log-linear schooling technology maps directly into the constant
schooling coefficient used in the empirical specification. Proposition~\ref{prop:schooling}
therefore predicts a negative treated-by-post-by-schooling coefficient.

The machine-substitution term makes the schooling prediction stronger than in a pure factor-augmentation model. With Cobb--Douglas production between $Z$ and $T$ ($\rho=0$), $\theta_Z$ is constant but $h$ falls as $M(A)$ rises, so the schooling semi-elasticity declines. With gross substitutability, $0<\rho<1$ for any admissible CES elasticity and $0<\theta_T<1$, implying $\rho\theta_T<1$: the outer share effect attenuates the decline but cannot reverse it. The maintained complementarity assumption remains essential for the positive early-career experience prediction below.

\subsection{The return to experience and technological obsolescence}\label{sec:experience-theory}

In the baseline technology, the log-wage return to an additional year of
potential experience is
\begin{equation}
r_X^T(A,S,X) \equiv \frac{\partial \ln w}{\partial X}=\theta_T(A,S,X)g(X), \qquad g(X)\equiv\frac{\tau'(X)}{\tau(X)}>0.
\label{eq:rxtransfer}
\end{equation}

The function $g(X)$ is the proportional increment to practice-based
knowledge generated by another year of experience. It is natural to
assume $g'(X)<0$, since learning through practice is fastest
near labour-market entry and diminishes as experience accumulates. From
equation~\eqref{eq:sharechanges},
\begin{equation}
\frac{\partial r_X^T}{\partial\ln A}=-\rho\theta_Z\theta_T\eta_A g(X)>0.
\label{eq:experiencechange}
\end{equation}

Thus, complementarity raises the return to the transferable tacit
component of experience at every point in the career, with the largest
increase tending to occur where the learning gradient $g(X)$ is greatest.
This mechanism corresponds most closely to the positive empirical effect
at three and five years of potential experience.

Accumulated experience need not be fully transferable across technologies. A model containing only transferable tacit knowledge cannot capture depreciation of routines tied to the previous task environment. We therefore allow a portion of accumulated experience to be specific to the pre-AI task environment. Let $q(X)$
denote the stock of routines, procedures, and technology-specific
know-how vulnerable to obsolescence, with $q'(X)>0$. Let
$\Omega(A)$ be the fraction of its wage value displaced by the new technology,
where $\Omega'(A)>0$. Log wages become
\begin{equation}
\ln w_{iot}=a_{ot}+\ln F\!\left(\phi(S_i)+M(A_{ot}),\tau(X_i)\right)-\Omega(A_{ot})q(X_i).
\label{eq:obsolescence}
\end{equation}

The return to experience is now
\begin{equation}
r_X(A,S,X)=\theta_T(A,S,X)g(X)-\Omega(A)q'(X).
\label{eq:rx}
\end{equation}

Consider an increase in AI capability from $A_0$ to $A_1$. The change in the
experience return is
\begin{equation}
\Delta r_X(X)=\left[\theta_T(A_1,S,X)-\theta_T(A_0,S,X)\right]g(X)-\left[\Omega(A_1)-\Omega(A_0)\right]q'(X).
\label{eq:deltarx}
\end{equation}

The first term is the complementarity effect, since practice-based
judgment becomes more valuable when codified content becomes abundant.
The second is the obsolescence effect, since experience tied to the
previous task technology loses value. If $g(X)$ declines with experience
and $q'(X)$ is constant or increasing, the complementarity effect
dominates early in the career and the obsolescence effect can dominate
later.

\begin{proposition}[Rotation of the experience profile]\label{prop:rotation}
Define
the complementarity gain as $C(X)=\left[\theta_T(A_1,S,X)-\theta_T(A_0,S,X)\right]g(X)$.
Suppose $q'(X)>0$ and the ratio $C(X)/q'(X)$ decreases
continuously with $X$. If complementarity exceeds obsolescence at low
experience but is smaller at high experience, there is a unique
crossover $X^*$ such that
\begin{equation}
\Delta r_X(X)>0 \text{ for } X<X^*, \qquad \Delta r_X(X)<0 \text{ for } X>X^*.
\label{eq:rotation}
\end{equation}
\end{proposition}

The wage-experience profile therefore rotates in favour of early-career
workers. The result does not interpret all senior experience as
obsolete. It says that the marginal year of measured potential
experience combines a transferable tacit component and a
technology-vintage component, and that their relative importance can
vary over the career.

The empirical regression approximates the experience profile with a
quadratic centred at fifteen years. Consequently, the estimated effect
of the shock on its slope is
\begin{equation}
\widehat{\Delta r_X}(X)=\delta_X+2(X-15)\delta_{XX}.
\label{eq:empiricalrotation}
\end{equation}

Equation~\eqref{eq:empiricalrotation} is the empirical counterpart of equation~\eqref{eq:deltarx}: the quadratic profile approximates the change in the experience slope as linear in experience. A
positive effect at low experience and a negative effect at high
experience require $\delta_{XX}<0$. The implied empirical rotation point is
\begin{equation}
X^*=15-\frac{\delta_X}{2\delta_{XX}},
\label{eq:crossover}
\end{equation}
which lies above fifteen years when $\delta_X>0$.
This is the theoretical counterpart of the rotation point reported in
the results. Because the quadratic restricts the change in the slope to be linear in experience, Section~\ref{sec:identification-results} re-estimates the profile with restricted cubic splines and a quartic; the rotation survives these more flexible forms (Appendix Table~\ref{tab:A_funcform} and Appendix Figure~\ref{fig:A_funcform}).

\subsection{Mapping the model into the empirical design}

Let $D_o$ indicate an occupation in the high-exposure group and $P_t$ the
period after the public release of ChatGPT. Write the change in
occupation-specific AI capability as
\begin{equation}
\ln A_{ot}=a_o+c_t+\kappa D_oP_t, \qquad \kappa>0,
\label{eq:aishock}
\end{equation}
where $a_o$ allows occupations to differ in baseline technology and $c_t$
captures common changes over time. Equation~\eqref{eq:aishock} does not require
low-exposure occupations to receive no AI improvement. It requires the
increase to be larger in the occupations classified as highly exposed.
The static difference-in-differences specification estimates the
wage-structure response to this differential increase.

For schooling, the population analogue of the triple-interaction
coefficient is
\begin{equation}
\delta_S=\left[r_S(T,\mathrm{post})-r_S(T,\mathrm{pre})\right]-\left[r_S(C,\mathrm{post})-r_S(C,\mathrm{pre})\right].
\label{eq:didS}
\end{equation}

Under Proposition~\ref{prop:schooling} and a larger rise in $A$ for treated occupations, $\delta_S<0$.
Occupation fixed effects absorb permanent wage-level
differences, treated-by-skill interactions allow the two groups to have
different returns before the shock, post-by-skill interactions absorb
common changes in those returns, and the treated-by-post term absorbs
differential changes in average wages. The identifying variation is
therefore the differential change in the schooling gradient within
occupations.

For experience, the corresponding estimand at $X$ years is
\begin{equation}
\delta_X(X)=\left[r_X(T,\mathrm{post};X)-r_X(T,\mathrm{pre};X)\right]-\left[r_X(C,\mathrm{post};X)-r_X(C,\mathrm{pre};X)\right].
\label{eq:didX}
\end{equation}

Proposition~\ref{prop:rotation} predicts $\delta_X(X)>0$ at low experience and
permits $\delta_X(X)<0$ at high experience. In the quadratic
empirical model, equation~\eqref{eq:didX} is represented by $\delta_X+2(X-15)\delta_{XX}$. The
event-study specification replaces the single change $\kappa$ with
year-specific changes. Its pre-release coefficients test whether treated
and control occupations were already experiencing different trends in
the schooling and experience gradients before generative AI became
widely available.

The model generates no necessary sign for the treated-by-post wage-level
coefficient. AI may raise output through augmentation, lower wages
through displacement or rents, or have different effects across
occupations. This indeterminacy is intentional, since the empirical
exercise tests a repricing of schooling and experience rather than a
general claim about average wage effects.

\subsection{Empirical predictions and scope}

The empirical exercise focuses on two implications.

\begin{itemize}
\item \textbf{Schooling premium.} The return to a year of schooling falls
differentially after November 2022 in occupations with high pre-existing AI exposure, so that $\delta_S<0$.

\item \textbf{Experience profile.} The return to experience rises differentially at
low levels of potential experience. If experience with the previous task
technology is sufficiently vulnerable to obsolescence, the effect
declines with experience and can become negative later in the career, so
that $\delta_{XX}<0$ and, under the conditions of Proposition~\ref{prop:rotation}, a
finite crossover $X^*$ exists.
\end{itemize}

These predictions rest on imperfect empirical proxies. Years of schooling capture codifiable knowledge only imperfectly, and potential experience measures opportunities to acquire practice-based knowledge rather than practice itself. Occupational exposure measures add another layer: they locate where AI capabilities overlap with job requirements, but say nothing on their own about whether codified and practice-based knowledge are complements. Reading the schooling estimate through the model therefore relies on an identifying assumption, not a tested one: that greater measured exposure corresponds to a larger post-2022 expansion in machine-supplied codified knowledge. Interpreting a rise in early-career experience returns as a complementarity effect requires the additional production assumption.

Because most schooling and
experience stocks are predetermined during the sample window, the
estimates identify changes in their prices rather than changes in
human-capital investment. The model also abstracts from aggregate
relative-supply effects across worker types. Such forces can move skill
prices, as in Katz and Murphy (1992); the empirical design removes them
only when they are common across exposure groups or reflect stable
pre-existing differences. A differential post-2022 change in worker
composition in exposed occupations remains a possible confound. Longer-run
responses involving education choices, AI-assisted learning, mentorship,
or intergenerational knowledge transmission require cohort or longitudinal
evidence and lie beyond the empirical exercise studied here.

\section{Related literature}\label{sec:literature}

Most of the related literature estimates the level of one of two
objects, the wage premium associated with an additional year of
schooling or the value of an additional year of labour-market
experience. We estimate changes in both jointly, guided by a framework
predicting that a technology which substitutes for codified knowledge
moves them in opposite directions. We first set out the conceptual
foundations: human-capital theory, tacit knowledge, and the task
framework. We then review how each return has been estimated, together
with the evidence that both behave as prices that respond to technology
and factor supply. We next turn to the small literature that studies the
two returns together and to the emerging evidence on generative AI, and
close with our contribution and its limits.

\subsection{Conceptual foundations}

The canonical framework is due to Becker (1964), Ben-Porath (1967) and
Mincer (1974). Ben-Porath provides the classic life-cycle account of
human-capital accumulation, while Mincer provides the empirical
expression we estimate in equation~\eqref{eq:static}. Our short-run model
takes existing schooling and experience as given rather than modelling
the investment decision itself. Two features of the Mincer equation
matter here. First, it is separable,
since the return to a year of schooling is assumed constant across
levels of experience. Heckman, Lochner and Todd (2006) show that this
separability, together with the stationarity assumptions needed to read
the schooling coefficient as an internal rate of return, is rejected in
United States data from the 1960s onward. Our specification relaxes
separability by construction, since the parameters of interest are
triple interactions. Second, the Mincer equation prices experience
without saying what experience produces. Rosen (1972) makes the point
explicit, since jobs are bundles of production and learning
opportunities, and Arrow (1962) supplies the production-side counterpart
in learning by doing.

Our framework rests on a distinction from outside labour economics.
Polanyi (1966) argued that a substantial part of what a skilled
practitioner knows cannot be stated in propositional form, since it is
acquired by doing, is bound to context, and is demonstrated rather than
described. Nelson and Winter (1982) carried the idea into economics
through organisational routines, and Szulanski (1996) documented the
internal stickiness firms encounter when transferring a practice they
already possess. The most direct labour-market evidence is Huckman and
Pisano (2006), who show that the performance of cardiac surgeons is
firm-specific, so that part of measured expertise resides in the
practice environment rather than in the transferable person. This
literature supplies the maintained assumption behind our second
prediction, that a component of what the experience profile prices is
absent from the documented record on which large language models are
trained.

The task framework of Autor, Levy and Murnane (2003), Acemoglu and Autor
(2011) and Acemoglu and Restrepo (2018, 2019) supplies the mechanism by
which a technology changes the value of a skill, displacing labour from
some tasks and raising the marginal product of labour in those that
remain. Katz and Murphy (1992) and Goldin and Katz (2008) supply the
demand-and-supply reading of the education premium as a race between
technology and the supply of schooled labour, the benchmark against
which any claim about a falling education premium has to be judged.

Closest to our prediction is Autor and Thompson (2025), who show that
automation raised wages where it removed inexpert tasks and lowered them
where it removed expert ones. We apply the same logic to a technology
specialised in codified content.

A very recent theoretical literature studies AI in economies where
knowledge is tacit or acquired on the job. Ide and Talamas (2025) show
that the gains from AI depend on its autonomy and fall unevenly across
workers by knowledge. Ide (2026) shows that entry-level automation can
slow the transmission of tacit knowledge from experts to novices, and
Cai (2026) that AI use raises short-run productivity but can slow skill
accumulation through learning by doing. Closest to our structure is
Acemoglu, Kong and Ozdaglar (2026), in which decisions combine shared
general knowledge with complementary context-specific knowledge. In
their model, as in Ide and Talamas (2025), AI supplies the tacit or
context-specific input, whereas in ours it supplies the codified one.
The two assumptions carry different implications for the experience
gradient, which our estimates can inform. These models concern
organisation, learning incentives and the long-run formation of skills;
our contribution relative to this group is a short-run, testable
statement about the relative prices of schooling and experience.

\subsection{Estimating the two returns}

The literature on the return to schooling is organised around a single
identification problem, that schooling is chosen, so that ordinary least
squares confounds the causal return with the earnings advantage of
whoever selects into more schooling. Card (1999, 2001) remain the
reference statements. Twin and sibling comparisons deliver returns of
roughly 9 to 13 percent, compulsory schooling reforms yield 11 to 18
percent (Oreopoulos 2006), admission-threshold discontinuities deliver
19 to 23 percent for marginal admits (Mountjoy 2026), and Patrinos and
Psacharopoulos (2026) report an average causal return near 10 percent
across 145 studies and 54 countries.

What we take from this literature is a boundary rather than a benchmark.
Every such design identifies the level of a return by finding exogenous
variation in the schooling decision, and none speaks to movements in the
price of schooling holding the decision fixed, which is our object. This
is why ability bias is not the first-order threat here. The classic
selection problem contaminates the level of the schooling coefficient.
For our difference-in-differences parameter, the relevant concern is
whether worker composition changed differentially in high-exposure
occupations after November 2022. Section~\ref{sec:descriptives} examines
observable composition; differential sorting on unobserved ability
remains a limitation.

Autor and Thompson (2025), discussed above, also clarify what these
wage gradients price: the relevant margin is the expertise required by
the remaining task bundle, not the presence of automation as such. This
is why we ask whether generative AI moves the schooling and experience
gradients in different directions, rather than whether it raises or
lowers wages.

The literature on experience is organised around a different problem,
since potential experience, actual experience, firm tenure and match
quality are jointly determined. The classic exchange runs from Altonji
and Shakotko (1987) to Topel (1991), with Altonji and Williams (2005)
settling on an intermediate magnitude. A second strand asks what
experience is specific to. Kambourov and Manovskii (2009) find that
occupational tenure carries most of the value, while Gathmann and
Schonberg (2010) refine this to task-specific human capital that
transfers in proportion to task distance. Both matter for us, since our
treatment is assigned at the occupation level and generative AI
displaces tasks rather than occupations.

The strand closest to our mechanism establishes that much of the
experience premium is produced by exposure to other people rather than
by the passage of time. Jarosch, Oberfield and Rossi-Hansberg (2021)
show that having more highly paid coworkers predicts subsequent wage
growth, and Herkenhoff, Lise, Menzio and Phillips (2024) estimate a
learning technology in teams. Arellano-Bover and Saltiel (2026) identify
firm-specific on-the-job learning premia from movers and show that they
persist after the worker leaves, so that what is learned is portable
even though it is produced in a specific place, while Lise and
Postel-Vinay (2020) add that multidimensional skills accumulate when
used and depreciate when not.

Deming and Noray (2020) show that faster changes in occupational skill
requirements are associated with flatter age--earnings profiles and
higher exit rates from STEM careers. Their model explains these patterns
as a race between on-the-job learning and skill obsolescence. We
examine how generative AI
may change both margins at once: complementary practice-based knowledge
can become more valuable early in the career, while experience tied to
older technologies and routines can lose value later on. Our estimates
trace this rotation of the experience gradient alongside the change in
the schooling gradient; they do not separately identify learning and
obsolescence rates.

Jeong, Kim and Manovskii (2015) establish that the return to experience
is a market price that responds to relative factor supply. Their model
shows that demographic changes can account for large movements in the
experience premium and for differences in wage-premium dynamics across
age and education groups. We instead estimate differential changes in
skill prices across occupational AI-exposure groups around the dated
release of ChatGPT in November 2022. Their findings also identify a
competing explanation: differential changes in the supply of experienced
workers could mimic technological repricing, making composition and
pre-trend checks central to our interpretation.

Humlum and Vestergaard (2025) link surveys of AI-chatbot adoption to
Danish administrative records and document widespread adoption and
changes in work tasks, but precisely estimated null effects on average
earnings and recorded hours. Their confidence intervals rule out
average effects larger than one percent in the period studied. Our
outcome is the relative price of schooling and experience, rather than
the average earnings effect of adoption. Changes in these gradients,
including offsetting effects across the experience distribution, can
therefore coexist with small average earnings effects.

A recent working paper by Absawy (2026) also combines CPS data with
occupational AI exposure and reports a rising education premium. Our
design differs in three respects: we estimate the schooling and
experience gradients jointly, allow the experience effect to vary over
the career, and treat top-coded earnings as censored under both Census
regimes. The last matters because top-coding is concentrated in exposed
occupations, where it affects 11.5 percent of observations against 0.7
percent elsewhere, and rises with education.\footnote{When the Census
Bureau replaced the fixed cap with a monthly-recomputed mean in April
2023, the values recorded for top-coded workers rose mechanically, so
treating them at face value can inflate the measured education premium
in exposed occupations in exactly the period after the release.}

\subsection{The two returns jointly, and the generative-AI shock}

The literature that estimates the two returns as one object is small,
and it is where this paper is most directly located. Deming (2025) is
the central reference. Following United States workers to age sixty, he
shows that the college wage premium roughly doubles over the life cycle,
from 27 percent at age 25 to 60 percent at age 55, and that its growth
is driven by sorting into professional, non-routine occupations with
steeper returns to tenure. Part of what schooling buys is therefore
access to jobs in which practice pays more, and our split of the sample by college credential
(Section~\ref{sec:heterogeneity-results}) is the natural test inside
that framework. Blundell et al.\ (2016) and Lagakos et al.\ (2018) find that returns to experience rise
with education, the structural counterpart of the baseline interactions
our specification allows.

Any claim that the education premium fell after 2022 has to be set
against the fact that it had already stopped rising (Autor, Goldin and
Katz 2020; Valletta 2018). This is why our design cannot rest on a time
series and requires the differential comparison, the pre-2022 placebo,
and the falsification reported in Section~\ref{sec:identification-results}. Our treatment definition
follows the occupational exposure literature. Felten, Raj and Seamans
(2021) build the AI Occupational Exposure index from the fifty-two
abilities in O*NET, deliberately agnostic about whether exposure implies
substitution or complementarity, while Eloundou, Manning, Mishkin and
Rock (2024) supply the alternative measure we use as a robustness check,
which embeds a substitution judgment the ability-based index avoids.

Most of the experimental and observational evidence on generative AI, from field and online experiments on worker productivity (Noy and Zhang 2023; Dell'Acqua et al.\ 2023; Brynjolfsson, Li and Raymond 2025b) to studies of adoption and employment, examines productivity, employment, hours, or earnings levels rather than the joint repricing of schooling and experience within exposed occupations. Our object is therefore different: whether the wage gradients associated with these two forms of human capital move around the technology shock. Our results are compatible
with the Danish nulls of Humlum and Vestergaard (2025), since a rotation
of the experience profile with an offsetting group-by-period term leaves
average earnings unchanged. A nascent literature asks whether AI changes
how much workers learn rather than how much they produce (Kim, Mester
and Sun 2026), and the evidence there is thin enough that our
early-career results are best presented as consistent with, rather than
as a test of, the formation channel.

\subsection{Contribution and limitations}

As the introduction emphasises, the contribution lies in the object estimated: the market prices attached to predetermined stocks of schooling and experience, treated as outcomes of a technology shock, with parallel trends required in skill prices rather than in wage levels. The joint movement of the two prices is what gives the exercise structure. A fall in the schooling gradient alone, or a rise in early-career experience returns alone, could reflect many changes in the labour market, such as shifting age composition or employer learning; the combination of a lower schooling return with an experience profile that rises at the bottom and falls at the top is more restrictive, and in the framework it maps into three distinct margins: substitution between human and machine codified knowledge, complementarity with transferable judgment, and obsolescence of technology-specific experience.

The main limitations follow from measurement and selection. Schooling and potential experience are coarse proxies: years of schooling do not distinguish fields or quality, while potential experience counts time since school rather than time spent practising the tasks exposed to AI. Selection into employment may also change after the shock. If young workers remaining in exposed occupations are positively selected, part of the early-career estimate could reflect composition rather than price. The concern is not hypothetical. Brynjolfsson, Chandar and Chen (2025a) document a relative decline in early-career employment in the most AI-exposed occupations with no comparable change for senior workers, and Ide (2026) reviews similar evidence from firm adoption and vacancy data. Finally, the schooling coefficient can embody signalling and employer learning as well as productive human capital. A technology that changes how quickly firms infer productivity could compress the schooling gradient without changing the productive content of schooling. The design differences out persistent differences in these mechanisms, but not a post-2022 change in how they operate. We therefore read the estimates as a repricing of measured schooling and potential experience, with the model offering one disciplined account rather than an exhaustive decomposition.

\section{Data and empirical strategy}\label{sec:data}

The empirical design compares labour market outcomes, specifically the
returns to education and experience, in occupations with high versus low
exposure to generative AI, before and after the public release of
ChatGPT in November 2022. This section describes the data used, the
exposure measurement and treatment definition, the descriptive
statistics, and the estimating equations.

\subsection{Data and sample}\label{sec:sample}

Our primary data source is the Current Population Survey Outgoing
Rotation Group (CPS-ORG). The CPS uses a 4-8-4 rotation sample design,
in which households are interviewed four consecutive months, leave the
sample for eight months, and then return for another four months. The
earnings supplement is administered only in the fourth and eighth
interviews, when households are asked about usual weekly earnings and
hours worked. Our data cover 2018 through June 2026, almost five years
of pre-treatment data and three and a half years of post-treatment data.
We aggregate observations into years relative to the timing of the
treatment shock, defined from November through October, so that no bin
mixes pre- and post-treatment months and monthly sampling noise averages
out.

Our worker sample consists of individuals between 16 and 64 years of age
who report a valid hourly wage. We measure hourly earnings using the
reported hourly wage for hourly workers and weekly earnings divided by
usual weekly hours worked for non-hourly workers. We consider an hourly
wage valid when it is non-missing and strictly positive, excludes wages
allocated by the BLS, and the individual reports hours worked. We also
impose a minimum hourly wage of \$3.625, half the federal minimum wage, for non-hourly workers,
following the conventions used in Katz and Murphy (1992). For hourly
workers we impose a lower floor of \$1, since their wage is reported
directly rather than derived from weekly earnings and hours, and since a
higher floor would exclude workers earning the federal tipped minimum of
\$2.13. These restrictions leave 655,919 observations in the dataset.

Top-coding in these data is concentrated almost entirely in the
treatment group. Censoring reaches 11.5 percent of observations in
high-exposure occupations, against 0.7 percent in low-exposure
occupations. Within each group censoring is unevenly distributed, rising
with education and higher among men, as Appendix Table~\ref{tab:A_censoring} shows.
Top-coding is therefore most frequent in the group where our parameters
are identified.

The censoring threshold is not constant over our sample period. Through
March 2023, the Census Bureau top-codes weekly earnings at a fixed value
of \$2,884.61, a nominal cap that has remained unchanged since 1998.
Beginning in April 2023 this cap is replaced by a value that Census
recomputes each month as the weighted mean of the top weekly earners.
The difference between the two regimes lies in the recorded value for
top-coded observations. Under the old regime, top-coded observations are
recorded at the threshold itself. Under the new regime, they are
recorded at the weighted mean of weekly earnings among the top three
percent of earners, recomputed each month, which by construction exceeds
the threshold. The new procedure was phased in for households entering the sample from January 2023, so it first affects earnings records in April 2023. Households interviewed before the change keep the old cap until they leave the sample, so the two regimes coexist until March 2024.

We assign each censored observation to its regime using the
month-in-sample variable. Under the static regime the lower bound is the
registered cap, common to all censored observations. Under the dynamic
regime the recorded value lies above the threshold and varies from month
to month, so we instead take the lower bound to be the highest
uncensored value recorded that month, computed separately for hourly and
non-hourly workers. Section~\ref{sec:identification-results} reports estimates in which the
dynamically top-coded observations are instead treated as point
observations at the value assigned by Census.

Educational attainment is recorded in sixteen categories, ranging from
less than first grade to a doctorate, which we convert into years of
schooling following Jaeger (1997). Potential experience is defined as
the maximum of age minus years of schooling minus six, or zero. We also
truncate it at 45 years, so that a few extreme values do not dominate
the curvature of the estimated experience profile.

Occupation is the unit at which the treatment is assigned. Through
December 2019, occupations are coded according to the 2010 Census
classification. From January 2020 onward, they are coded according to
the 2018 classification. We map the 2018 taxonomy codes to the 2010
counterpart using the crosswalk published by IPUMS-CPS, and then convert
all of them into the 2010 Standard Occupational Classification System
(SOC10) using the census occupation code list, the taxonomy on which our
principal exposure measure is built.

\subsection{Occupational AI exposure and the arrival of generative AI}

Our primary exposure measure is the AI Occupational Exposure (AIOE) index constructed by Felten, Raj and Seamans (2021). Our design interacts this pre-existing occupational exposure measure with the arrival of generative AI after November 2022.
AIOE matches ten AI applications to the fifty-two abilities described in
O*NET and then aggregates these links to occupations using the
prevalence and importance of each ability within each occupation. An
occupation receives a higher score when the abilities it requires are
those that AI can perform. The index was also constructed before the
release of ChatGPT, so it cannot have been fitted to labour market
outcomes observed after the shock.

As the authors state, this measure is agnostic about the consequences of
exposure. It records how much the abilities an occupation requires
overlap with what AI can do, and it stays silent on whether AI is a
complement to or a substitute for labour. The index identifies which
occupations are exposed, while what exposure does to the returns is a
prediction of the framework in Section~2 rather than a property of the
index. The index is also invariant over time, so all the time variation
in our design comes from the shock rather than from the definition of
the treatment.

As noted in the previous section, AIOE is built for detailed SOC 2010
occupations, while the census codes recorded in our panel combine
detailed occupations with broader groupings. We therefore aggregate the
index to the panel' s occupational codes using an
employment-weighted average, with employment weights taken from the 2018
Occupational Employment Statistics and therefore fixed before the shock.
This results in 64.7 percent of observations matched to a detailed
occupation, 33.6 percent to a broader group, and 1.7 percent receiving
no score. The unscored cells correspond to groups for which no exposure information is available and to residual categories. These observations have no exposure quartile and are therefore excluded from the estimation sample.

We define treatment by the extreme quartiles of the exposure
distribution. Workers whose occupations fall in the top quartile of
AIOE, and whose required abilities are sensitive to AI applications,
form the treatment group, while those in the bottom quartile, whose
required abilities are least affected by AI applications, form the
control group. We exclude the two middle quartiles because they are
neither clearly exposed nor clearly unexposed. Section~\ref{sec:scope} reports
estimates using alternative partitions of the exposure distribution. The
treated group includes occupations such as software developers,
financial managers, and lawyers, while the control group includes
occupations such as truck drivers, construction labourers, and waiters.
The quartile cutoffs are computed once over the entire panel, so no
occupation moves between groups over time. Computing the cutoffs only from pre-release data, January 2018 to October 2022, moves three occupations that lie at the boundaries and leaves the estimates essentially unchanged (Appendix Table~\ref{tab:A_precutoffs}).

We date the shock to November 2022, when ChatGPT was released to the public and large language models became widely known. This provides a discrete public-release date for a technology whose workplace adoption then proceeded rapidly. Measured from each technology's first mass-market launch, adoption of generative AI was extraordinarily fast: overall adoption outpaced that of the personal computer and the internet, and adoption at work matched that of the personal computer (Bick, Blandin and Deming 2024). Treated observations are those of top-quartile workers from the release onward. Diffusion still took time even at this pace, so the effect need not be uniform across post-release years. The event studies in Section~\ref{sec:results-eventstudy} follow it year by year.

Another measure of exposure in this literature is the GPTs are GPTs
index of Eloundou, Manning, Mishkin and Rock (2024). It differs from
AIOE in what it measures, not only in how it is built. The measure asks
whether a large language model can substantially reduce the human time a
task requires while keeping the same quality, so it embeds a judgment
about substitution that AIOE deliberately avoids. Estimates and further
explanation about this measure are in Section~\ref{sec:scope}.

Figure~\ref{fig:exposure_groups} compares the two measures across major occupation groups. They
broadly agree on who is exposed. Management, business, legal, and
computer occupations rank at the top under both, and manual and service
occupations at the bottom. Since the indices share neither data nor
method, one built from the abilities an occupation requires and the
other from judgments about its tasks, the agreement is not mechanical,
and the treatment assignment does not depend on the construction choices
of either team. The measures are not interchangeable, however, and they
disagree most in a few groups such as education. Accordingly, we present
estimates under each measure and identify the results that are sensitive
to the choice of measure. Section~\ref{sec:scope} re-estimates the model under the Eloundou et al.\ measure and under two of the ten applications that make up AIOE, image generation and language modelling (Appendix Table~\ref{tab:A_aioe_apps} lists all ten), to ask whether different AI capabilities reprice different margins.

\subsection{Descriptive statistics}\label{sec:descriptives}

Table~\ref{tab:sumstats} summarises the sample we use to estimate the returns to skills. Columns (1) and (2) report weighted averages of individual characteristics for the control and treated groups in the pre-treatment period, while columns (3) and (4) report them for the post-treatment period. Panel A summarises wages, the outcome variable, and panel B the composition of the two groups. Levels differ markedly between the groups, as one would expect if high-exposure occupations are concentrated in analytical and credential-intensive work (Felten, Raj and Seamans 2021). Table~\ref{tab:occupations} lists the largest occupations in each group. Before the shock, wages are 65 log points higher among the treated than among controls, treated workers have 3.6 more years of education, and they are eight times as likely to hold a college degree. What remains fixed after the shock is not only the relative size of the two groups, whose share of total employment moves from 24.1 to 24.5 percent, but also their composition, since at the occupation level the changes in average education and experience are uncorrelated with exposure (Appendix Figure~\ref{fig:A_composition}). The main observable movements are a 0.8 percentage point increase in the share of college-educated workers and a 1.4 point decline in the share of workers under thirty in the treated group, relative to controls (Table~\ref{tab:sumstats}). Both movements pre-date the shock, and the differential change in schooling is not statistically significant, while the treated workforce ages by 0.3 years relative to the control group (Appendix Table~\ref{tab:A_composition}). Relative to controls, the share married rises by 2.5 percentage points and the share paid hourly falls by 1.1 points (Appendix Table~\ref{tab:A_composition}). Both enter the individual controls, and hourly pay also enters the variance model.

The last two rows of panel B report two further features of the treated
group that our estimation must address. First, treated workers are
rarely paid by the hour (26 percent, against 82 percent among controls),
which matters because wages are more dispersed around their predicted
value among salaried workers than among hourly ones. We therefore let
the error variance differ between the two, as described in Section~\ref{sec:estimation}. Second, a substantially higher fraction of the treated group is
top coded (12.4 versus 0.5 percent in the pre period), which
concentrates censoring precisely where we want to estimate effects and
motivates the interval estimator we present in Section~\ref{sec:estimation}.

Figure~\ref{fig:exposure_schooling} shows why we focus on the return to education as one of the
skill prices of interest. The graph plots occupational cells by their
exposure score against the average years of education (panel A) and
average potential experience (panel B) of workers in the pre period. We
observe a pronounced gradient in education associated with exposure
across the whole occupation distribution. In fact, the overlap between
the two groups on education is almost non-existent. Control occupations
have workers with 10 to 14 years of schooling on average, while treated
cells have workers with 13 to 19 years. By contrast, there is no
variation in experience associated with exposure across occupations, and
the two groups are virtually identical in experience. Sorting
occupations by exposure therefore sorts them by the schooled, codifiable
component of human capital while leaving the practice-based component
balanced. Finally, the returns to education and experience themselves
differ substantially across the two groups. In the pre period, the
return to an additional year of education is 7.5 log points for the treated group, compared with 1.7 for the control group (Appendix Table~\ref{tab:A_returns_group}). Similarly, the log wage gap between workers with 25 versus 5 years
of experience is 26.8 log points in the treated sample, compared with
11.9 in the control sample (Figure~\ref{fig:profile_pre}).

Figure~\ref{fig:exp_by_occ} shows that this difference is not confined to the extremes of
the exposure distribution. When we estimate a separate wage equation
within each occupational cell, the log wage differential between 25 and
5 years of experience rises steadily with exposure: it averages 20 log points in bottom-quartile occupations and 34 in top-quartile ones, with a
slope of 6.0 (s.e. 0.9). The return to experience rises with exposure
across the whole occupational distribution, not only between the groups
we compare. Exposed occupations thus have both the highest return to schooling and
the highest return to experience. Both returns are high in the
occupations the technology affects, so the shock can move them in
opposite directions, the pattern the model predicts and Section~5 tests.

\subsection{Specification}\label{sec:specification}

We estimate the effect of exposure on the returns to schooling and
experience with the following wage equation,

\begin{align}
\ln w_{it}={}&\beta_S S_i+\beta_X x_i+\beta_{XX}x_i^2
+\gamma_S(D_oS_i)+\gamma_X(D_ox_i)+\gamma_{XX}(D_ox_i^2)\notag\\
&+\pi_S(P_tS_i)+\pi_X(P_tx_i)+\pi_{XX}(P_tx_i^2)\notag\\
&+\delta_S(D_oP_tS_i)+\delta_X(D_oP_tx_i)+\delta_{XX}(D_oP_tx_i^2)+\delta_0(D_oP_t)\notag\\
&+C_i'\theta+\mu_m+\nu_s+\lambda_o+\varepsilon_{it}.
\label{eq:static}
\end{align}

where $S_i$ is years of education, $X_i$ is potential experience, and $x_i\equiv X_i-15$ is centred potential experience. $D_o$ indicates a treated occupation, $P_t$ indicates the post-November 2022 period, and $C_i$ is a vector of individual controls. The vector of individual controls includes indicators for sex, race and ethnicity, marital status, union membership, public-sector employment, metropolitan residence, full- or part-time status, citizenship, and pay type (salaried, hourly, or hourly with tips, overtime or commissions). The terms $\mu_m$, $\nu_s$, and $\lambda_o$ are month, state, and occupation fixed effects, respectively. Since treatment is assigned at occupation level and is time-invariant, $D_o$ is absorbed by the occupation fixed effects, while $P_t$ is absorbed by the month fixed effects.

The parameters of interest are $\delta_S$, $\delta_X$, and $\delta_{XX}$. What distinguishes these
from standard difference-in-differences (DiD) estimates is that the
outcome to which the two differences are applied is the return to skill
rather than the wage itself. Notice from the estimating equation that
the return to education is $\beta_S$ for the control group in the pre period,
$\beta_S$ + $\pi_S$ for the control group in the post period, $\beta_S$ + $\gamma_S$ for the
treated group in the pre period, and $\beta_S$ + $\gamma_S$ + $\pi_S$ + $\delta_S$ for the treated
group in the post period. Thus, $\delta_S$ is a DiD estimator of the change in
the return to education for treated versus control occupations, the
empirical test of the first prediction in Section~2.

Since the experience profile is quadratic, the impact of the shock on
the return to experience depends on the point at which this impact is
evaluated. It equals $\delta_X+2(X-15)\delta_{XX}$ at $X$ years of experience, and we
present the results evaluated at three, five, fifteen, and twenty-five
years. This is the empirical counterpart of the second prediction of
Section~2, and its shape over $X$ is what allows the shock to raise the
return to experience for some workers while lowering it for others.

The two blocks of double-interaction terms play the role of the group
and time indicators of a DiD, and not including them would be equivalent
to estimating a DiD without group and period indicators. The $\gamma$ terms
allow the returns to differ between exposed and unexposed occupations at
baseline, and the $\pi$ terms allow them to change over time for all
workers. Without controlling for them, the parameter $\delta_S$ would absorb any
pre-existing difference in returns between the two groups, or any change
in returns common to both.

\subsection{Event study}\label{sec:eventstudy}

The previous specification imposes a single break at the date of the
shock. To understand the dynamic effects, as well as to check for
parallel trends before the shock, we replace the post indicator with a
set of indicators for years relative to the release. We define $B_{tb}$ as an indicator for the year $b$ relative to the release,
where years run from November to October and $b=0$ corresponds to
November 2022 through October 2023. The estimating equation becomes

\begin{align}
\ln w_{it}={}&\beta_S S_i+\beta_X x_i+\beta_{XX}x_i^2
+\gamma_S(D_oS_i)+\gamma_X(D_ox_i)+\gamma_{XX}(D_ox_i^2)\notag\\
&+\sum_{b\neq -1}\!\left[\pi_{Sb}(B_{tb}S_i)+\pi_{Xb}(B_{tb}x_i)+\pi_{XXb}(B_{tb}x_i^2)\right]\notag\\
&+\sum_{b\neq -1}\!\left[\delta_{Sb}(D_oB_{tb}S_i)+\delta_{Xb}(D_oB_{tb}x_i)
+\delta_{XXb}(D_oB_{tb}x_i^2)+\delta_{0b}(D_oB_{tb})\right]\notag\\
&+C_i'\theta+\mu_m+\nu_s+\lambda_o+\varepsilon_{it}.
\label{eq:eventstudy}
\end{align}

where $B_{tb}$ indicates relative year $b$ and both sums run over all years except $b=-1$, the year immediately preceding the release, which is the omitted category.

The coefficients $\delta_{Sb}$ capture the change in the differential return to
education in treated versus control occupations in year b, relative to
the year before the release. This means the event study coefficients
estimate the differential returns rather than the differential level of
wages, in the same units as the static specification. The effect on the
return to experience again depends on the point at which it is
evaluated, and we report $\delta_{Xb}+2(X-15)\delta_{XXb}$ at the same points used in
the previous section.

As in the static specification we also include the double interactions
that allow the returns to schooling and experience to vary freely across
years and to differ between the two groups at baseline. These terms play
the same role as the $\gamma$ and $\pi$ terms above, and their omission would make
the event study coefficients capture the common fluctuations of the
returns across years.

We assess the validity of the identifying assumption by studying the
estimated coefficients for the years before the shock. We report joint
tests that all pre-shock coefficients are zero, separately for the
schooling and experience margins, together with other specification
tests. These tests are presented with the estimates in Section~\ref{sec:results-eventstudy}.

\subsection{Identification, estimation, and inference}\label{sec:estimation}

Identification of $\delta_S$, $\delta_X$, and $\delta_{XX}$ relies on parallel
trends in the return to education and in the experience profile between
treated and control occupations in the absence of the shock, conditional
on the fixed effects and individual controls. This is the analogue of the
standard DiD assumption, applied to skill prices rather than wage levels:
it does not require parallel trends in average wages, and it is weaker
than requiring parallel wage trends for every schooling--experience cell.
Shocks that move the level of wages, whether common to all workers or
specific to the treated group, are captured by the month fixed effects and
$\delta_0$ and do not affect the coefficients of interest. The assumption
therefore rests on the absence of any event, other than the introduction
of generative AI, that changes the return to education or experience
differentially in treated occupations after November 2022, and on the
composition of the two groups remaining stable, which
Section~\ref{sec:descriptives} examines.

We estimate the wage equation directly on the censored data, without
imputing top-coded wages. Uncensored wages enter the likelihood as points
and censored wages as an interval between the lower bound of their regime
and positive infinity, and we estimate the resulting interval-censored
model by maximum likelihood (Stewart 1983). The censoring documented in
Section~\ref{sec:sample} is what makes this choice appropriate. The usual
alternative in work on the U.S. wage structure assigns top-coded earnings
a fixed multiple of the cap (Lemieux 2006; Autor, Katz and Kearney 2008).
Here, top-coding falls almost entirely on treated occupations, so any
imputation rule would apply almost exclusively to the observations that
identify $\delta_S$, $\delta_X$, and $\delta_{XX}$, and the rule would then
determine the education gradient in the upper tail, where much of the variation that identifies $\delta_S$ lies.

Consistency rests on two parametric assumptions. The first is that,
conditional on covariates, log wages are normally distributed with the
conditional mean in equation~\eqref{eq:static}, so that the upper tail of
the normal distribution accounts for the censored observations. This is
the assumption behind the censored-regression treatment of top-coded
earnings in administrative wage data, where Tobit models estimated within
cells are standard (Dustmann, Ludsteck and Schonberg 2009; Card, Heining
and Kline 2013). Two features of our setting limit how much it can matter.
With 5.9 percent of the sample censored, and 11.5 percent of the treated
group, normality only has to describe a small share of the data, and the
biases from non-normality are small at these censoring rates (Arabmazar
and Schmidt 1982). The censoring bounds are also observed rather than
estimated, since under both top-coding regimes the bound of each interval
is constructed from the data as described in Section~\ref{sec:sample}.
Section~\ref{sec:identification-results} reports a check in which the
dynamically top-coded observations enter as points at the values the
Census assigns. Because the dynamic regime begins in April 2023, the check
affects few observations in the release year; in later post years it moves
the schooling estimate toward zero, for reasons discussed there, while
leaving the experience rotation unchanged. Section~\ref{sec:identification-results} also imposes a common censoring rule across the full sample, which leaves the estimates close to the baseline (Table~\ref{tab:censoring}).

The second assumption concerns the error variance. Unlike least squares,
where heteroskedasticity affects only standard errors, unmodelled
heteroskedasticity makes the censored maximum likelihood estimator
inconsistent, with a bias that grows with the censoring rate (Arabmazar
and Schmidt 1981). Because censoring falls almost entirely on treated
workers, unmodelled heteroskedasticity would bias the two groups
differently, and that difference would contaminate the parameters of
interest. We therefore estimate the variance jointly with the mean,
letting the log standard deviation of the error depend on year, an
indicator for hourly-paid workers, and years of schooling, the dimensions
along which residual wage dispersion varies most in these data. This
follows the logic of the cell-specific Tobit models cited above, which let
the variance differ by year, education and age, but models it
parametrically within a single likelihood.
Section~\ref{sec:identification-results} shows that richer variance
specifications leave the results essentially unchanged.

All specifications are weighted by the CPS earnings weights, so that the
estimates describe the population of U.S. wage and salary earners rather
than the sample. Section~\ref{sec:identification-results} reports
unweighted estimates, which are statistically indistinguishable from
these. Standard errors are clustered at the occupation level, the level at
which the treatment is assigned (Bertrand, Duflo and Mullainathan 2004;
Abadie, Athey, Imbens and Wooldridge 2023), with 253 occupation clusters.
Evidence on the identifying assumption, including joint tests that the
pre-shock event-study coefficients are zero and sensitivity to violations
of parallel trends (Rambachan and Roth 2023), is reported with the results
in Section~\ref{sec:results-eventstudy}.

\section{Results}\label{sec:results}

Two facts organise the results. After the release of ChatGPT, the schooling gradient falls in exposed occupations and the experience gradient rotates, rising early in the career and falling later on. We begin with these estimates and what they imply about the nature of the knowledge being repriced, then turn to timing, identification and robustness, and heterogeneity and scope.

\subsection{The two skill prices}\label{sec:schooling-results}

Table~\ref{tab:main} reports estimates of equation~\eqref{eq:static}. Column 5 is our baseline specification. The coefficient $\delta_S$ measures the differential post-release change in the return to a year of schooling in treated relative to control occupations.

A year of schooling pays 0.55 log points less in treated occupations after November 2022 than it does under the counterfactual evolution of the control group (s.e. 0.19). Before the shock, the return to a year of schooling was 7.5 log points in treated occupations, more than four times the corresponding return in control occupations. The estimate therefore represents a decline of roughly seven percent relative to its own pre-shock level. Appendix Table~\ref{tab:A_returns_group} decomposes the estimate by group and period, and Appendix Figure~\ref{fig:A_returns_group_time} plots the return to schooling by group and year. The return to schooling in control occupations is essentially unchanged after the release, at 1.7 log points in both periods. So the difference-in-differences estimate reflects almost entirely the decline in treated occupations, from 7.5 to 6.9 log points, a change of -0.54 (s.e. 0.15).

The estimate becomes larger and more precise once occupation fixed effects enter. This matters for interpretation. The comparison is not simply between highly educated occupations and less educated ones; it is between workers with different schooling inside the same occupation, before and after the shock. In the language of the model, the result is consistent with machine-supplied codified knowledge reducing the scarcity value of the worker-supplied component. It does not imply that educated workers became less productive in levels. It implies that the wage differential associated with an additional year of schooling became smaller in the occupations most exposed to the technology.

\label{sec:experience-results}The experience result is not a level shift. Because equation~\eqref{eq:static} is quadratic in experience, the post-release effect on the slope of the wage-experience profile varies with $X$. Table~\ref{tab:main} evaluates that effect at selected points, while Figure~\ref{fig:profile_effect} plots it over the full support.

At three years of potential experience, the return to an additional year rises by 0.26 log points in treated occupations (s.e. 0.10); at five years, by 0.23 (s.e. 0.09). The effect is statistically indistinguishable from zero at fifteen years and equals -0.12 log points at twenty-five (s.e. 0.04). Under the baseline quadratic, the implied crossover is 18.1 years of potential experience (s.e. 2.1). In other words, the wage-experience profile rotates in favour of less-experienced workers rather than shifting uniformly. Section~\ref{sec:identification-results} shows that flexible experience profiles preserve the rotation but not a uniquely pinned-down crossover age.

This shape is consistent with Proposition~\ref{prop:rotation}, although the reduced-form estimates do not separately identify its two mechanisms. One interpretation is that complementarity with practice-based judgment dominates at low experience, where the learning gradient from practice is steep, while technology-specific obsolescence matters more later in the career. The result does not imply that senior experience is generally obsolete; it indicates that the marginal year of measured experience can contain both transferable and technology-vintage components whose relative values change after the shock.

The pooled wage effect is correspondingly concentrated among highly schooled workers. For a worker with sixteen years of schooling and fifteen years of experience in a treated occupation, the estimated wage change relative to the counterfactual path is -3.2 log points (s.e. 0.8). Workers with less schooling lose considerably less. This level effect is not itself a theoretical prediction; it is the wage implication of the estimated changes in skill prices combined with the treated-by-post intercept.

\paragraph{One kind of knowledge, or two?} The three estimates also
discriminate between accounts of what AI does to knowledge. Recent models
of AI in the knowledge economy treat knowledge as a single stock
(Section~\ref{sec:literature}). Mapping that stock to experience, we can
read off what two of them imply for the margins we estimate. In Ide and
Talamas (2025), knowledge is a scalar capacity to solve problems, and AI
that acts only as a copilot benefits the least knowledgeable workers
most. Read as experience, this flattens the whole profile: it matches the
late-career decline but runs against the early-career rise. In Ide
(2026), tacit knowledge is learned by doing entry-level work alongside
experts, and entry-level automation raises the income of current experts
relative to novices. This matches the early-career rise, but the model
has only two career stages, so it is silent on the profile at twenty-five
years and on a crossover. Ide (2026) also describes formal education as
supplying codifiable knowledge, close to our distinction, but the model
has a single stock of knowledge and attaches no separate price to its
codified part. Neither framework therefore speaks to the return to
schooling.

Each one-dimensional view thus matches part of the pattern. A framework
with two kinds of knowledge, priced separately, reproduces all three
movements: machine-supplied codified knowledge lowers the return to
schooling, complementarity raises the value of practice early in the
career, and obsolescence lowers it later on. These readings of the two
alternative models are ours, with knowledge mapped to potential
experience; neither paper makes predictions about Mincer gradients.
Sections~\ref{sec:results-eventstudy} and~\ref{sec:identification-results}
assess the credibility of each margin, and especially of the late-career
decline.

\subsection{Dynamics and pre-trends}\label{sec:results-eventstudy}

The static estimates would be difficult to interpret if the skill-price gaps were already moving before November 2022. Figures~\ref{fig:es_schooling} and~\ref{fig:es_experience} therefore replace the single post indicator with event-time indicators whose years run from November to October; Appendix Table~\ref{tab:A_es_full} reports the coefficients and the joint pre-trend tests.

For schooling, the four pre-release coefficients are small, ranging from -0.12 to 0.10 log points, and the joint test that they are zero yields $p=0.90$. The break appears in the first twelve months after the release, when the differential return to schooling falls by 0.91 log points (s.e. 0.33). The post-release coefficients are not monotone, so we do not interpret their year-to-year variation as a diffusion path; the robust feature is a flat pre-period followed by a negative post-release average. Rambachan--Roth calculations (Appendix Table~\ref{tab:A_honestdid}) show that the release-year effect survives deviations from parallel trends as large as the largest pre-period deviation.

For experience, the two ends of the profile begin moving in opposite directions after the release. At three years, every post-release coefficient is positive and three of four are statistically significant. At twenty-five years, the coefficients are negative throughout the post-period and significant in every year after the release year. This simultaneous divergence is the empirical counterpart of the rotation in Proposition~\ref{prop:rotation} and is hard to reconcile with a simple level shock to wages.

The late-career event study is the weaker of the two margins. The joint pre-trend test at twenty-five years gives $p=0.06$, and one coefficient four years before the release is individually significant. We therefore place more weight on the early-career increase and on the full-profile rotation than on a literal causal interpretation of every late-career point estimate.

\subsection{Identification and robustness}\label{sec:identification-results}

The most plausible threats to the design concern timing, censoring, permanent differences in occupation-specific skill prices, and the contemporaneous technology-sector downturn.

\paragraph{Pandemic and placebo timing.} The pandemic occupies a substantial part of the pre-period and affected office-intensive occupations differently from manual and service work. Yet the pandemic-year event-study coefficients for both schooling and experience are small and statistically insignificant. More directly, Table~\ref{tab:falsification} re-dates the shock to January 2020 and estimates the model only on 2018--2021 data. The schooling effect is -0.02 log points ($p=0.94$), the experience effects are null, and the pre-trend tests do not reject. The design does not mechanically generate a break around a major labour-market disruption.

A second falsification contrasts the second and bottom exposure quartiles, neither of which is treated. It produces a schooling estimate similar in magnitude to the headline result but fails the pre-trend test ($p=0.03$). It also produces no experience rotation: the effect is 0.03 log points at three years and -0.01 at twenty-five, neither statistically significant. The raw series show why (Appendix Figure~\ref{fig:A_q2q1}): the schooling-return gap between these two low-exposure groups was already narrowing before ChatGPT and does not break at the release. The exercise shows that differences in schooling composition can generate a large schooling estimate when the pre-trend condition fails, but not the joint pattern of a falling schooling gradient and a rotating experience profile; the extreme-quartile comparison used in the baseline passes the pre-trend test and displays both.

\paragraph{Top-coding.} Censoring is concentrated in treated occupations and the Census procedure changes in April 2023, making top-coding the most consequential measurement issue in the paper. Treating the dynamically top-coded observations as point values at the Census-assigned monthly means is an intentionally harsh check. It gives every top-coded worker in a month the same wage, so after April 2023 wages in the upper tail no longer vary with schooling, and the schooling estimate falls from -0.55 to -0.09 (Table~\ref{tab:estimation}, column 4; Appendix Table~\ref{tab:A_pointdata} reports the corresponding event study): the procedure removes within-tail variation that identifies the schooling gradient. The experience rotation remains.

A cleaner test imposes a common artificial censoring rule throughout the sample. We apply the pre-2023 weekly cap of \$2,884.61 in every month while preserving the censoring information already present in the CPS: observations that were originally top-coded remain censored, with their lower bound set to the smaller of the original Census threshold and the common cap; observations that were originally uncensored are newly censored only when they exceed the common cap. Re-estimating the same interval-censored model gives a schooling effect of -0.53 log points (s.e. 0.18), while the return to experience rises by 0.25 at three years (s.e. 0.10) and falls by 0.12 at twenty-five (s.e. 0.04), with a crossover of 18.0 years (Table~\ref{tab:censoring}, column 1). More aggressive common weekly caps of \$2,500 and \$2,000 yield schooling effects of -0.40 and -0.45 log points, with the same qualitative rotation (columns 2 and 3). The event study under the common cap is also close to the baseline, and its pre-trend tests give $p=0.87$ for schooling and $p=0.55$ and $p=0.06$ for experience at three and twenty-five years. Because the artificial cap is imposed symmetrically before and after April 2023 without discarding the information contained in the original top-coding, the headline pattern cannot be generated mechanically by the change in Census top-coding regime.

\paragraph{Occupation-specific skill slopes.} Occupation fixed effects absorb permanent wage-level differences but not permanent differences in the way occupations price schooling and experience. We therefore allow each occupation its own pre-shock coefficients on schooling, centred experience, and centred experience squared. Under this specification the schooling effect is -0.58 log points (s.e. 0.17); the return to experience is +0.26 at three years (s.e. 0.09) and -0.11 at twenty-five (s.e. 0.04), with a crossover at 18.3 years (Table~\ref{tab:censoring}, column 4). The pre-trend tests of the corresponding event study give $p=0.90$, $p=0.47$ and $p=0.08$. The result is therefore not inherited from fixed cross-occupation differences in baseline Mincer slopes.

\paragraph{Pre-release exposure cutoffs.} The quartile cutoffs that define the treatment and control groups are computed over the full panel, so they partly reflect post-release observations. Computing them on data from January 2018 to October 2022 only reclassifies three occupations at the boundaries and leaves the results essentially unchanged: the schooling effect is -0.56 log points (s.e. 0.19), the return to experience rises by 0.26 at three years and falls by 0.13 at twenty-five, the crossover is 17.8 years, and the pre-trend tests give $p=0.87$, $p=0.64$ and $p=0.07$ (Appendix Table~\ref{tab:A_precutoffs}).

\paragraph{Technology-sector downturn.} The release of ChatGPT coincided with large layoffs in technology firms. Excluding computer and mathematical occupations removes 15 percent of the treated group but leaves the schooling decline at -0.49 log points and the experience rotation essentially unchanged (Table~\ref{tab:droptech}). Excluding the information sector yields a schooling estimate of -0.59. A sector-specific downturn would also have to change the schooling and experience gradients inside occupations, rather than simply wage levels, to reproduce the estimates.

\paragraph{Functional form of the experience profile.} The quadratic implies that the effect on the slope of the wage--experience profile is linear in experience, so the rotation could reflect that restriction rather than the data. Murphy and Welch (1990) argue that a quartic describes experience profiles better than a quadratic. We therefore re-estimate the model with a quartic and with a restricted cubic spline in potential experience, with knots at the percentiles recommended by Harrell (2001), and interact each with treatment, post, and treatment $\times$ post. Appendix Table~\ref{tab:A_funcform} reports the effect on the return to experience at selected points, and Appendix Figure~\ref{fig:A_funcform} plots it over the full profile. The broad pattern is the same in every specification: positive early in the career, close to zero around mid-career, and negative later on. The quadratic lies inside the confidence band of the five-knot spline at every level of experience. The flexible specifications are less precise at particular points, especially early in the career, and they place the onset of the late-career decline somewhat later than the quadratic. At twenty-five years the five-knot spline is close to zero, while by thirty to thirty-five years the flexible profiles are clearly negative. The rotation is therefore not an artefact of the quadratic, but the exact crossover point is not tightly pinned down. We treat the 18.1-year baseline crossover as a useful summary of the quadratic profile rather than as a structural threshold.

\paragraph{Weights and variance model.} The results are not sensitive to the remaining estimation choices. Removing CPS earnings weights leaves the parameters statistically indistinguishable from the baseline, and richer variance specifications move the schooling estimate by less than four percent (Table~\ref{tab:estimation}, columns 1 to 3). These results, together with the common-censoring and occupation-slope exercises above, make the two main findings less dependent on the particular estimation choices used in the baseline.

None of these checks overturns the central pattern. The schooling result survives both a common censoring rule and occupation-specific skill slopes, while the experience rotation is even more stable across specifications. The main qualification remains the weaker pre-trend evidence at the high-experience end of the profile.

\subsection{Heterogeneity and scope}\label{sec:heterogeneity-results}

The repricing is concentrated among workers for whom codified knowledge was especially important. Appendix Table~\ref{tab:A_credential} splits the sample between workers with and without a college credential (associate degree or higher). Among credentialed workers, the return to experience rises by 0.43 log points at three years (s.e. 0.16) and falls by 0.12 at twenty-five (s.e. 0.07). Among workers without a credential, the early-career gain is smaller and the late-career decline is not statistically significant. Because the schooling result also has its largest wage implications for the highly educated, the two movements affect the same broad group rather than unrelated parts of the workforce.

This split is informative for the mechanism. Within the credentialed group, where everyone has already crossed an education threshold, wage differences by early-career experience become steeper after the shock. That is consistent with a technology that makes codified content easier to obtain while leaving practice-based judgment relatively scarce. The larger late-career decline among credentialed workers is also consistent with obsolescence of experience accumulated in applying knowledge that the new technology now supplies more directly, although age and credential status cannot be separated perfectly with potential-experience data.

We also decompose AIOE into its image-generation and language-modelling components (Table~\ref{tab:measures}, columns 2 and 3; Figures~\ref{fig:es_school_comp} and~\ref{fig:es_exp3_comp}). These exercises are exploratory rather than central to identification because the components rank occupations differently and are strongly correlated across occupations. Image-generation exposure produces a larger schooling estimate, while language-model exposure reproduces more of the experience rotation. The latter result comes with an important caveat: the pre-trend test for the late-career effect under the language-model measure rejects at the five-percent level ($p=0.03$). We therefore treat the decomposition as suggestive evidence that different AI capabilities may reprice different margins of human capital, not as a separate causal result.

\label{sec:scope}Turning to scope, the dose-response exercise points in the same direction as the baseline. Table~\ref{tab:dose} contrasts each exposure quintile with the bottom quintile. The total wage effect for a college-educated worker rises in magnitude from 0.9 percent in the second quintile to 3.0 percent in the top. The early-career experience effect is close to zero in the lower quintiles and reaches 0.25 log points in the top quintile. The schooling coefficient is less monotone in the middle of the distribution because those contrasts inherit the pre-existing trend identified by the low-exposure falsification; the top-versus-bottom contrast is both the largest and the cleanest in pre-trends.

Replacing AIOE with the task-based measure of Eloundou et al.\ (Table~\ref{tab:measures}, column 1) produces a schooling decline of 0.42 log points ($p=0.02$), close to the baseline. The early-career rotation is much weaker under that measure, although the late-career decline remains. The difference is informative about scope: exposure measures that classify a different set of occupations need not recover the same experience mechanism. The schooling result is comparatively robust to the choice of exposure metric; the positive early-career experience result is more specific to the occupations selected by AIOE.

The remaining scope limitation is measurement. Years of schooling treat an engineering degree and a literature degree as the same quantity of human capital, while potential experience counts years since school rather than years spent practising the tasks that AI affects. The estimates therefore describe changes in the prices of generic schooling and potential experience. They do not imply that every field of study or every form of senior expertise has been devalued in the same way.

\section{Conclusion}\label{sec:conclusion}

Generative AI can change the wage structure before it changes the stock of human capital. In the short window studied here, workers' schooling and accumulated experience are largely predetermined. What moves is their price. In occupations most exposed to AI, the return to schooling falls after the release of ChatGPT, while the return to experience rotates: it rises early in the career and falls later on.

The codified/practice distinction gives these movements a natural interpretation. Machine-supplied codified knowledge competes directly with the codified component of worker human capital, reducing the scarcity value associated with schooling. At the same time, an abundance of codified content can raise the value of complementary judgment learned through practice. The latter effect is strongest where learning is fastest, early in the career. For more experienced workers, however, part of accumulated know-how is tied to routines and technologies that the new system changes, creating an obsolescence margin. The mid-career crossover is the reduced-form expression of these opposing forces; its exact location depends on the functional form used to trace the experience profile.

The findings are not a forecast of long-run returns to education or experience. Schooling choices, curricula, AI-assisted learning, mentoring, occupational sorting, and the organisation of entry-level work will all adjust. Indeed, the longer-run question may be whether workers can continue to acquire the practice-based knowledge that becomes more valuable when AI automates parts of the learning process itself. The results here establish a prior step: the arrival of generative AI is already associated with a measurable repricing of existing human capital inside exposed occupations.

\clearpage
\nocite{*}
\bibliographystyle{apalike}
\bibliography{references}

\clearpage
\appendix
\section*{Tables}
\addcontentsline{toc}{section}{Tables}
{\small

\begin{table}[H]
\centering
\caption{Summary statistics by exposure group and period}
\label{tab:sumstats}
\begin{tabular}{@{}
  >{\raggedright\arraybackslash}p{(\columnwidth - 8\tabcolsep) * \real{0.4154}}
  >{\raggedright\arraybackslash}p{(\columnwidth - 8\tabcolsep) * \real{0.1462}}
  >{\raggedright\arraybackslash}p{(\columnwidth - 8\tabcolsep) * \real{0.1462}}
  >{\raggedright\arraybackslash}p{(\columnwidth - 8\tabcolsep) * \real{0.1462}}
  >{\raggedright\arraybackslash}p{(\columnwidth - 8\tabcolsep) * \real{0.1462}}@{}}
\toprule\noalign{}
\begin{minipage}[b]{\linewidth}\raggedright
\end{minipage} & \begin{minipage}[b]{\linewidth}\raggedright
\textbf{Control}
\end{minipage} & \begin{minipage}[b]{\linewidth}\raggedright
\textbf{Treated}
\end{minipage} & \begin{minipage}[b]{\linewidth}\raggedright
\textbf{Control}
\end{minipage} & \begin{minipage}[b]{\linewidth}\raggedright
\textbf{Treated}
\end{minipage} \\
\midrule\noalign{}
& (1) Pre & (2) Pre & (3) Post & (4) Post \\
\emph{\textbf{Panel A. Wages}} & & & & \\
Log hourly wage & 2.795 & 3.447 & 3.008 & 3.695 \\
\emph{\textbf{Panel B. Composition}} & & & & \\
Years of schooling & 12.17 & 15.81 & 12.26 & 15.89 \\
Share with college degree & 0.087 & 0.709 & 0.100 & 0.730 \\
Potential experience & 19.96 & 19.73 & 19.63 & 19.61 \\
Share under 30 & 0.329 & 0.190 & 0.333 & 0.180 \\
Share female & 0.287 & 0.531 & 0.291 & 0.529 \\
Share of total employment & 0.259 & 0.241 & 0.249 & 0.245 \\
Share paid hourly & 0.820 & 0.262 & 0.817 & 0.247 \\
Share top-coded & 0.005 & 0.124 & 0.009 & 0.104 \\
Observations & 98,174 & 93,325 & 63,566 & 64,205 \\
\bottomrule\noalign{}
\end{tabular}

\vspace{0.5em}
\begin{minipage}{\linewidth}
Notes: Weighted means using ORG earnings weights. The pre-treatment
period runs from January 2018 to October 2022 and the post-treatment
period from November 2022 to June 2026. Wages are nominal. Treated
(control) occupations are those in the top (bottom) quartile of the AIOE
distribution.
\end{minipage}
\end{table}
\clearpage

\begin{table}[H]
\centering
\caption{The largest occupations in the top and bottom exposure quartiles}
\label{tab:occupations}
\begin{tabular}{@{}
  >{\raggedright\arraybackslash}p{(\columnwidth - 6\tabcolsep) * \real{0.5077}}
  >{\raggedright\arraybackslash}p{(\columnwidth - 6\tabcolsep) * \real{0.1641}}
  >{\raggedright\arraybackslash}p{(\columnwidth - 6\tabcolsep) * \real{0.1641}}
  >{\raggedright\arraybackslash}p{(\columnwidth - 6\tabcolsep) * \real{0.1641}}@{}}
\toprule\noalign{}
\begin{minipage}[b]{\linewidth}\raggedright
\textbf{Occupation}
\end{minipage} & \begin{minipage}[b]{\linewidth}\raggedright
\textbf{AIOE}
\end{minipage} & \begin{minipage}[b]{\linewidth}\raggedright
\textbf{Years of schooling}
\end{minipage} & \begin{minipage}[b]{\linewidth}\raggedright
\textbf{Share of quartile employment}
\end{minipage} \\
\midrule\noalign{}
\emph{\textbf{Panel A. Treated occupations (top quartile, AIOE between
1.06 and 1.53)}} & & & \\
Software developers, applications and systems software & 1.23 & 16.4 &
6.8\% \\
Secretaries and administrative assistants & 1.11 & 14.0 & 6.3\% \\
Accountants and auditors & 1.48 & 16.2 & 4.9\% \\
Financial managers & 1.45 & 15.5 & 4.0\% \\
Postsecondary teachers & 1.13 & 18.4 & 3.7\% \\
Chief executives & 1.33 & 16.3 & 3.6\% \\
Marketing and sales managers & 1.29 & 15.6 & 3.6\% \\
Secondary school teachers & 1.14 & 17.2 & 3.3\% \\
\emph{\textbf{Panel B. Control occupations (bottom quartile, AIOE
between -2.29 and -0.72)}} & & & \\
Driver/sales workers and truck drivers & -1.09 & 12.5 & 7.9\% \\
Laborers and freight, stock, and material movers, hand & -1.71 & 12.3 &
5.9\% \\
Janitors and building cleaners & -1.44 & 11.8 & 5.4\% \\
Nursing, psychiatric, and home health aides & -0.82 & 12.8 & 5.1\% \\
Waiters and waitresses & -0.95 & 12.7 & 4.9\% \\
Stock clerks and order fillers & -0.79 & 12.5 & 4.8\% \\
Construction laborers & -1.62 & 11.2 & 4.0\% \\
Production workers, all other & -1.29 & 12.3 & 3.1\% \\
\bottomrule\noalign{}
\end{tabular}

\vspace{0.5em}
\begin{minipage}{\linewidth}
Notes: The eight largest occupations by employment in the top and bottom
quartiles of the AIOE distribution, over the pre-treatment period. AIOE is the
employment-weighted mean of the index within the occupation; shares are
of total employment in the quartile. Occupation titles follow the 2010
Census classification. The two middle quartiles span AIOE values from
-0.71 to 0.23 and from 0.23 to 1.06; quartile cutoffs are computed over
the full sample, unweighted.
\end{minipage}
\end{table}
\clearpage

\begin{table}[H]
\centering
\setstretch{1.0}
\caption{Changes in skill returns by occupational AI exposure after ChatGPT}
\label{tab:main}
\begin{tabular}{@{}
  >{\raggedright\arraybackslash}p{(\columnwidth - 10\tabcolsep) * \real{0.4155}}
  >{\raggedright\arraybackslash}p{(\columnwidth - 10\tabcolsep) * \real{0.1169}}
  >{\raggedright\arraybackslash}p{(\columnwidth - 10\tabcolsep) * \real{0.1169}}
  >{\raggedright\arraybackslash}p{(\columnwidth - 10\tabcolsep) * \real{0.1169}}
  >{\raggedright\arraybackslash}p{(\columnwidth - 10\tabcolsep) * \real{0.1169}}
  >{\raggedright\arraybackslash}p{(\columnwidth - 10\tabcolsep) * \real{0.1169}}@{}}
\toprule\noalign{}
\begin{minipage}[b]{\linewidth}\raggedright
\end{minipage} & \begin{minipage}[b]{\linewidth}\raggedright
\textbf{(1)}
\end{minipage} & \begin{minipage}[b]{\linewidth}\raggedright
\textbf{(2)}
\end{minipage} & \begin{minipage}[b]{\linewidth}\raggedright
\textbf{(3)}
\end{minipage} & \begin{minipage}[b]{\linewidth}\raggedright
\textbf{(4)}
\end{minipage} & \begin{minipage}[b]{\linewidth}\raggedright
\textbf{(5)}
\end{minipage} \\
\midrule\noalign{}
\emph{\textbf{Panel A. Return to schooling}} & & & & & \\
Change in the return to a year of schooling & -0.49** & -0.42* & -0.34 &
-0.59*** & -0.55*** \\
& (0.25) & (0.24) & (0.23) & (0.22) & (0.19) \\
Return in treated occupations, pre-shock & 11.94*** & 11.89*** &
11.47*** & 9.29*** & 7.48*** \\
& (0.92) & (0.91) & (0.88) & (0.53) & (0.45) \\
\emph{\textbf{Panel B. Return to a year of experience}} & & & & & \\
At 3 years of experience & 0.16 & 0.19 & 0.20 & 0.24** & 0.26*** \\
& (0.15) & (0.15) & (0.14) & (0.10) & (0.10) \\
At 5 years of experience & 0.14 & 0.17 & 0.18 & 0.21** & 0.23** \\
& (0.14) & (0.13) & (0.13) & (0.09) & (0.09) \\
At 15 years of experience & 0.03 & 0.04 & 0.06 & 0.05 & 0.05 \\
& (0.07) & (0.07) & (0.07) & (0.05) & (0.05) \\
At 25 years of experience & -0.08* & -0.08* & -0.05 & -0.10** & -0.12*** \\
& (0.05) & (0.05) & (0.05) & (0.04) & (0.04) \\
Experience-profile rotation point (years) & 17.64 & 18.63 & 20.36 &
18.42 & 18.11 \\
& (5.37) & (4.46) & (4.11) & (2.48) & (2.14) \\
\emph{\textbf{Panel C. Level effect}} & & & & & \\
Total effect, worker with 16 years of schooling & -1.65 & -1.29 & -1.08
& -2.69*** & -3.15*** \\
& (1.34) & (1.28) & (1.25) & (0.94) & (0.81) \\
Month fixed effects & No & Yes & Yes & Yes & Yes \\
State fixed effects & No & No & Yes & Yes & Yes \\
Occupation fixed effects & No & No & No & Yes & Yes \\
Individual controls & No & No & No & No & Yes \\
Observations & 315,930 & 315,930 & 315,930 & 315,930 & 315,930 \\
Occupation clusters & 253 & 253 & 253 & 253 & 253 \\
\bottomrule\noalign{}
\end{tabular}

\vspace{0.5em}
\begin{minipage}{\linewidth}
Notes: Interval-censored maximum likelihood estimates of equation (22),
weighted by ORG earnings weights, with standard errors clustered by
occupation in parentheses. Each column adds a block of controls to the
previous one; column 5 is the specification used throughout the paper.
The sample is held fixed across columns at the observations with
non-missing controls. The variance model is the same in every column.
Because the treatment is time-invariant within occupation and the post
indicator is a function of the month, a treated-group indicator is
entered explicitly in columns 1 to 3 and a post indicator in column 1,
where the corresponding fixed effects do not absorb them. The return to
experience is evaluated as the derivative of the estimated profile; the
rotation point is the experience level at which it changes sign, with a
delta-method standard error. * p\textless0.10, ** p\textless0.05, ***
p\textless0.01. Coefficients and standard errors are multiplied by 100
and read as log points.
\end{minipage}
\end{table}
\clearpage

\begin{table}[H]
\centering
\caption{Falsification tests}
\label{tab:falsification}
\begin{tabular}{@{}
  >{\raggedright\arraybackslash}p{(\columnwidth - 4\tabcolsep) * \real{0.6000}}
  >{\raggedright\arraybackslash}p{(\columnwidth - 4\tabcolsep) * \real{0.2000}}
  >{\raggedright\arraybackslash}p{(\columnwidth - 4\tabcolsep) * \real{0.2000}}@{}}
\toprule\noalign{}
\begin{minipage}[b]{\linewidth}\raggedright
\end{minipage} & \begin{minipage}[b]{\linewidth}\raggedright
\textbf{(1)}
\end{minipage} & \begin{minipage}[b]{\linewidth}\raggedright
\textbf{(2)}
\end{minipage} \\
\midrule\noalign{}
Change in the return to schooling & -0.02 & -0.49*** \\
& (0.22) & (0.18) \\
Return to experience, at 3 years & -0.06 & 0.03 \\
& (0.12) & (0.07) \\
at 25 years & 0.06 & -0.01 \\
& (0.04) & (0.03) \\
Pre-trend p, return to schooling & {[}0.92{]} & {[}0.03{]} \\
Pre-trend p, return to experience at 3 years & {[}0.37{]} &
{[}0.95{]} \\
Pre-trend p, return to experience at 25 years & {[}0.91{]} &
{[}0.38{]} \\
Placebo release date (January 2020) & Yes & No \\
Contrast: second vs bottom quartile & No & Yes \\
\bottomrule\noalign{}
\end{tabular}

\vspace{0.5em}
\begin{minipage}{\linewidth}
Notes: Column 1 assigns a placebo release date of January 2020 and
estimates the preferred specification on data from January 2018 to
December 2021, before the actual release. Column 2 contrasts the second
and bottom exposure quartiles, neither of which is treated; its
schooling estimate is close to the main effect but its pre-trend test
for schooling rejects, indicating a trend already present before the
shock rather than a break at the release. Pre-trend rows report, in
brackets, p-values of joint tests that the pre-shock event-study
coefficients are zero; in column 1 the placebo window leaves a single
pre-shock coefficient per margin. Standard errors clustered by
occupation in parentheses. * p\textless0.10, ** p\textless0.05, ***
p\textless0.01. Coefficients and standard errors are multiplied by 100
and read as log points.
\end{minipage}
\end{table}
\clearpage

\begin{table}[H]
\centering
\caption{Robustness to estimation choices}
\label{tab:estimation}
\begin{tabular}{@{}
  >{\raggedright\arraybackslash}p{(\columnwidth - 8\tabcolsep) * \real{0.4154}}
  >{\raggedright\arraybackslash}p{(\columnwidth - 8\tabcolsep) * \real{0.1462}}
  >{\raggedright\arraybackslash}p{(\columnwidth - 8\tabcolsep) * \real{0.1462}}
  >{\raggedright\arraybackslash}p{(\columnwidth - 8\tabcolsep) * \real{0.1462}}
  >{\raggedright\arraybackslash}p{(\columnwidth - 8\tabcolsep) * \real{0.1462}}@{}}
\toprule\noalign{}
\begin{minipage}[b]{\linewidth}\raggedright
\end{minipage} & \begin{minipage}[b]{\linewidth}\raggedright
\textbf{(1)}
\end{minipage} & \begin{minipage}[b]{\linewidth}\raggedright
\textbf{(2)}
\end{minipage} & \begin{minipage}[b]{\linewidth}\raggedright
\textbf{(3)}
\end{minipage} & \begin{minipage}[b]{\linewidth}\raggedright
\textbf{(4)}
\end{minipage} \\
\midrule\noalign{}
Change in the return to schooling & -0.44** & -0.57*** & -0.55*** &
-0.09 \\
& (0.18) & (0.21) & (0.20) & (0.20) \\
Return to experience, at 3 years & 0.21*** & 0.26*** & 0.26*** & 0.40*** \\
& (0.08) & (0.10) & (0.10) & (0.10) \\
at 25 years & -0.11*** & -0.12*** & -0.12*** & -0.10*** \\
& (0.03) & (0.04) & (0.04) & (0.03) \\
Pre-trend p, return to schooling & {[}0.57{]} & {[}0.85{]} & {[}0.90{]}
& {[}0.91{]} \\
Pre-trend p, return to experience at 3 years & {[}0.41{]} & {[}0.63{]} &
{[}0.55{]} & {[}0.59{]} \\
Pre-trend p, return to experience at 25 years & {[}0.12{]} & {[}0.06{]}
& {[}0.06{]} & {[}0.07{]} \\
CPS earnings weights & No & Yes & Yes & Yes \\
Variance model & Baseline & Extended & Baseline + regime & Baseline \\
Dynamic top-codes as point data & No & No & No & Yes \\
\bottomrule\noalign{}
\end{tabular}

\vspace{0.5em}
\begin{minipage}{\linewidth}
Notes: Each column re-estimates the preferred specification of Table~\ref{tab:main},
column 5, with the changes indicated at the foot of the table. The
baseline variance model lets the log standard deviation of the error
depend on year, an indicator for hourly-paid workers, and years of
schooling. Column 2 extends it with sex, the treatment group indicator,
potential experience and its square, and year-by-group interactions;
column 3 adds indicators for the top-coding regime to the baseline.
Column 4 enters dynamically top-coded wages as point observations at
their Census-assigned values rather than as censored intervals; because
the Census assigns a single value per month, the mean of the top three
percent, this variant also compresses the variance within the upper
tail. Pre-trend rows report, in brackets, p-values of joint tests that the pre-shock event-study coefficients are zero, for the return to schooling and for the return to experience at three and twenty-five years. Standard errors clustered by
occupation in parentheses. * p\textless0.10, ** p\textless0.05, ***
p\textless0.01. Coefficients and standard errors are multiplied by 100
and read as log points.
\end{minipage}
\end{table}
\clearpage

\begin{table}[H]
\centering
\caption{Common censoring rule and occupation-specific skill slopes}
\label{tab:censoring}
\begin{tabular}{@{}
  >{\raggedright\arraybackslash}p{(\columnwidth - 8\tabcolsep) * \real{0.4154}}
  >{\raggedright\arraybackslash}p{(\columnwidth - 8\tabcolsep) * \real{0.1462}}
  >{\raggedright\arraybackslash}p{(\columnwidth - 8\tabcolsep) * \real{0.1462}}
  >{\raggedright\arraybackslash}p{(\columnwidth - 8\tabcolsep) * \real{0.1462}}
  >{\raggedright\arraybackslash}p{(\columnwidth - 8\tabcolsep) * \real{0.1462}}@{}}
\toprule\noalign{}
\begin{minipage}[b]{\linewidth}\raggedright
\end{minipage} & \begin{minipage}[b]{\linewidth}\raggedright
\textbf{(1)}
\end{minipage} & \begin{minipage}[b]{\linewidth}\raggedright
\textbf{(2)}
\end{minipage} & \begin{minipage}[b]{\linewidth}\raggedright
\textbf{(3)}
\end{minipage} & \begin{minipage}[b]{\linewidth}\raggedright
\textbf{(4)}
\end{minipage} \\
\midrule\noalign{}
Change in the return to schooling & -0.53*** & -0.40** & -0.45** &
-0.58*** \\
& (0.18) & (0.17) & (0.17) & (0.17) \\
Return to experience, at 3 years & 0.25*** & 0.27*** & 0.23** & 0.26*** \\
& (0.10) & (0.10) & (0.10) & (0.09) \\
at 5 years & 0.21** & 0.24*** & 0.20** & 0.22*** \\
& (0.09) & (0.09) & (0.09) & (0.08) \\
at 15 years & 0.05 & 0.06 & 0.05 & 0.06 \\
& (0.04) & (0.04) & (0.04) & (0.04) \\
at 25 years & -0.12*** & -0.11*** & -0.11*** & -0.11*** \\
& (0.04) & (0.04) & (0.04) & (0.04) \\
Experience-profile rotation point (years) & 17.96 & 18.58 & 18.07 & 18.33 \\
& (2.21) & (1.93) & (2.31) & (1.94) \\
Total effect, 16 years of schooling & -3.33*** & -2.78*** & -3.08*** &
-3.17*** \\
& (0.80) & (0.79) & (0.77) & (0.81) \\
Pre-trend p, return to schooling & {[}0.87{]} & & & {[}0.90{]} \\
Pre-trend p, return to experience at 3 years & {[}0.55{]} & & &
{[}0.47{]} \\
Pre-trend p, return to experience at 25 years & {[}0.06{]} & & &
{[}0.08{]} \\
Common weekly cap & \$2,884.61 & \$2,500 & \$2,000 & No \\
Occupation-specific skill slopes & No & No & No & Yes \\
Observations & 315,930 & 315,930 & 315,930 & 315,930 \\
\bottomrule\noalign{}
\end{tabular}

\vspace{0.5em}
\begin{minipage}{\linewidth}
Notes: Each column re-estimates the preferred specification of Table~\ref{tab:main},
column 5. Columns 1 to 3 impose the indicated weekly cap in every month
of 2018--2026; for hourly workers the cap is the smaller of \$99.99 and
the weekly cap divided by usual hours. Observations that were originally
top-coded remain censored, with lower bound equal to the smaller of the
original Census threshold and the common cap; originally uncensored
observations are censored when they exceed the common cap. Column 4
keeps the original censoring and replaces the treated-group
interactions with schooling, experience, and experience squared by
occupation-specific interactions. Pre-trend rows report, in brackets,
p-values of joint tests that the pre-shock coefficients of the
corresponding event study are zero; the event study was estimated for
columns 1 and 4 only. The bottom rows of Table~\ref{tab:main} apply otherwise.
Standard errors clustered by occupation in parentheses. *
p\textless0.10, ** p\textless0.05, *** p\textless0.01. Coefficients and
standard errors are multiplied by 100 and read as log points.
\end{minipage}
\end{table}
\clearpage

\begin{table}[H]
\centering
\caption{Excluding the technology sector}
\label{tab:droptech}
\begin{tabular}{@{}
  >{\raggedright\arraybackslash}p{(\columnwidth - 4\tabcolsep) * \real{0.6000}}
  >{\raggedright\arraybackslash}p{(\columnwidth - 4\tabcolsep) * \real{0.2000}}
  >{\raggedright\arraybackslash}p{(\columnwidth - 4\tabcolsep) * \real{0.2000}}@{}}
\toprule\noalign{}
\begin{minipage}[b]{\linewidth}\raggedright
\end{minipage} & \begin{minipage}[b]{\linewidth}\raggedright
\textbf{(1)}
\end{minipage} & \begin{minipage}[b]{\linewidth}\raggedright
\textbf{(2)}
\end{minipage} \\
\midrule\noalign{}
Change in the return to schooling & -0.49** & -0.59*** \\
& (0.20) & (0.19) \\
Return to experience, at 3 years & 0.28*** & 0.26*** \\
& (0.10) & (0.10) \\
at 25 years & -0.11*** & -0.12*** \\
& (0.04) & (0.04) \\
Total effect, 16 years of schooling & -3.21*** & -3.38*** \\
& (0.85) & (0.82) \\
Excluded group & Computer and math occupations & Information sector \\
Share of the treated group excluded & 14.8\% & 3.0\% \\
Observations & 297,857 & 313,567 \\
\bottomrule\noalign{}
\end{tabular}

\vspace{0.5em}
\begin{minipage}{\linewidth}
Notes: Each column re-estimates the preferred specification excluding
the indicated group from the sample. Column 1 drops workers in computer
and mathematical occupations (SOC major group 15); column 2 drops all
workers in the information sector. Shares excluded are
employment-weighted shares of the treated group. The bottom rows of
Table~\ref{tab:main} apply otherwise. Standard errors clustered by occupation in
parentheses. * p\textless0.10, ** p\textless0.05, *** p\textless0.01.
Coefficients and standard errors are multiplied by 100 and read as log
points.
\end{minipage}
\end{table}
\clearpage

\begin{table}[H]
\centering
\caption{Alternative exposure measures}
\label{tab:measures}
\begin{tabular}{@{}
  >{\raggedright\arraybackslash}p{(\columnwidth - 6\tabcolsep) * \real{0.5077}}
  >{\raggedright\arraybackslash}p{(\columnwidth - 6\tabcolsep) * \real{0.1641}}
  >{\raggedright\arraybackslash}p{(\columnwidth - 6\tabcolsep) * \real{0.1641}}
  >{\raggedright\arraybackslash}p{(\columnwidth - 6\tabcolsep) * \real{0.1641}}@{}}
\toprule\noalign{}
\begin{minipage}[b]{\linewidth}\raggedright
\end{minipage} & \begin{minipage}[b]{\linewidth}\raggedright
\textbf{(1)}
\end{minipage} & \begin{minipage}[b]{\linewidth}\raggedright
\textbf{(2)}
\end{minipage} & \begin{minipage}[b]{\linewidth}\raggedright
\textbf{(3)}
\end{minipage} \\
\midrule\noalign{}
Change in the return to schooling & -0.42** & -0.70*** & -0.14 \\
& (0.18) & (0.18) & (0.16) \\
Return to experience, at 3 years & 0.06 & 0.14 & 0.33*** \\
& (0.08) & (0.09) & (0.10) \\
at 25 years & -0.09*** & -0.17*** & -0.10*** \\
& (0.03) & (0.04) & (0.03) \\
Pre-trend p, return to schooling & {[}0.89{]} & {[}0.93{]} &
{[}0.23{]} \\
Pre-trend p, return to experience at 3 years & {[}0.91{]} & {[}0.67{]} &
{[}0.86{]} \\
Pre-trend p, return to experience at 25 years & {[}0.16{]} & {[}0.41{]}
& {[}0.03{]} \\
GPT index (Eloundou et al.) & Yes & No & No \\
Image-generation component of AIOE & No & Yes & No \\
Language-modeling component of AIOE & No & No & Yes \\
\bottomrule\noalign{}
\end{tabular}

\vspace{0.5em}
\begin{minipage}{\linewidth}
Notes: Each column re-estimates the preferred specification of Table~\ref{tab:main},
column 5, replacing the AIOE index with the measure indicated at the
foot of the table; in each case the treatment contrasts the top and
bottom quartiles of the corresponding index. Pre-trend rows report, in
brackets, p-values of joint tests that the pre-shock event-study
coefficients are zero, for the return to schooling and for the return to
experience evaluated at three and at twenty-five years. Standard errors
clustered by occupation in parentheses. * p\textless0.10, **
p\textless0.05, *** p\textless0.01. Coefficients and standard errors are
multiplied by 100 and read as log points.
\end{minipage}
\end{table}
\clearpage

\begin{table}[H]
\centering
\caption{Dose response, each exposure quintile against the bottom quintile}
\label{tab:dose}
\begin{tabular}{@{}
  >{\raggedright\arraybackslash}p{(\columnwidth - 8\tabcolsep) * \real{0.4154}}
  >{\raggedright\arraybackslash}p{(\columnwidth - 8\tabcolsep) * \real{0.1462}}
  >{\raggedright\arraybackslash}p{(\columnwidth - 8\tabcolsep) * \real{0.1462}}
  >{\raggedright\arraybackslash}p{(\columnwidth - 8\tabcolsep) * \real{0.1462}}
  >{\raggedright\arraybackslash}p{(\columnwidth - 8\tabcolsep) * \real{0.1462}}@{}}
\toprule\noalign{}
\begin{minipage}[b]{\linewidth}\raggedright
\end{minipage} & \begin{minipage}[b]{\linewidth}\raggedright
\textbf{(1)}
\end{minipage} & \begin{minipage}[b]{\linewidth}\raggedright
\textbf{(2)}
\end{minipage} & \begin{minipage}[b]{\linewidth}\raggedright
\textbf{(3)}
\end{minipage} & \begin{minipage}[b]{\linewidth}\raggedright
\textbf{(4)}
\end{minipage} \\
\midrule\noalign{}
Change in the return to schooling & -0.47** & -0.48** & -0.30 &
-0.61*** \\
& (0.21) & (0.22) & (0.24) & (0.19) \\
Return to experience, at 3 years & 0.09 & 0.08 & 0.15* & 0.25** \\
& (0.08) & (0.08) & (0.09) & (0.11) \\
at 25 years & 0.01 & -0.09** & -0.04 & -0.13*** \\
& (0.03) & (0.04) & (0.03) & (0.04) \\
Total effect, 16 years of schooling & -0.92 & -1.85* & -2.64*** &
-2.99*** \\
& (1.11) & (0.96) & (0.98) & (0.86) \\
Treated group & Quintile 2 & Quintile 3 & Quintile 4 & Quintile 5 \\
Observations & 257,747 & 255,133 & 255,330 & 251,915 \\
\bottomrule\noalign{}
\end{tabular}

\vspace{0.5em}
\begin{minipage}{\linewidth}
Notes: Each column re-estimates the preferred specification with the
treatment defined as the indicated quintile of the AIOE distribution
against the bottom quintile; quintiles are constructed over the full
panel before sample restrictions, mirroring the construction of the
official quartiles. The bottom rows of Table~\ref{tab:main} apply otherwise. Standard
errors clustered by occupation in parentheses. * p\textless0.10, **
p\textless0.05, *** p\textless0.01. Coefficients and standard errors are
multiplied by 100 and read as log points.
\end{minipage}
\end{table}
\clearpage

\section*{Figures}

\begin{figure}[H]
\centering
\includegraphics[width=0.88\textwidth]{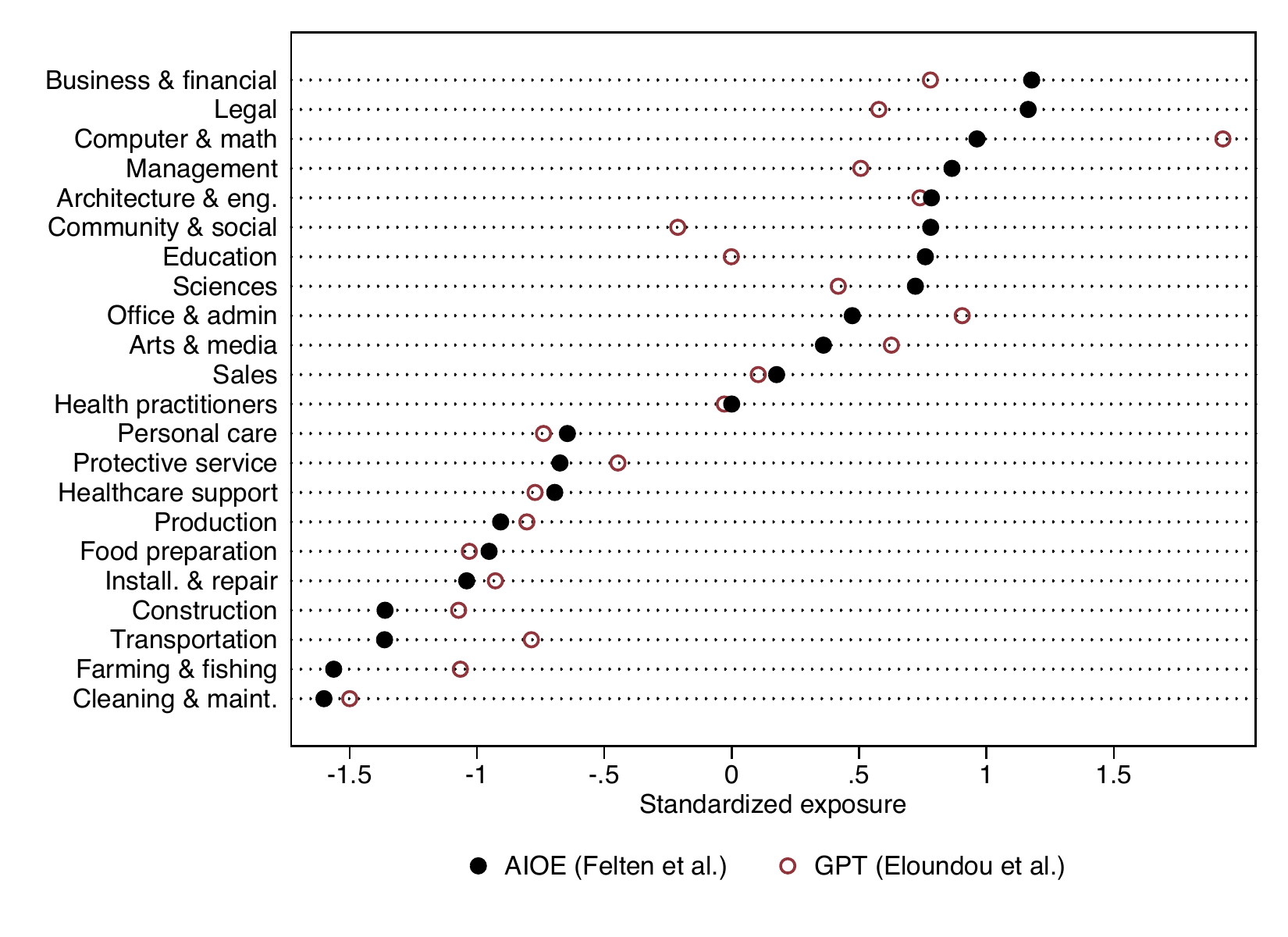}
\caption{Occupational AI exposure by major occupation group}
\label{fig:exposure_groups}
\begin{minipage}{0.92\textwidth}\footnotesize \textit{Notes:} Employment-weighted means of each standardized exposure measure within major occupation groups, computed over the pre-treatment period. Both measures are standardized to mean zero and unit standard deviation across occupational cells. Groups are ordered by AIOE.\end{minipage}
\end{figure}
\clearpage

\begin{figure}[H]
\centering
\includegraphics[width=0.98\textwidth]{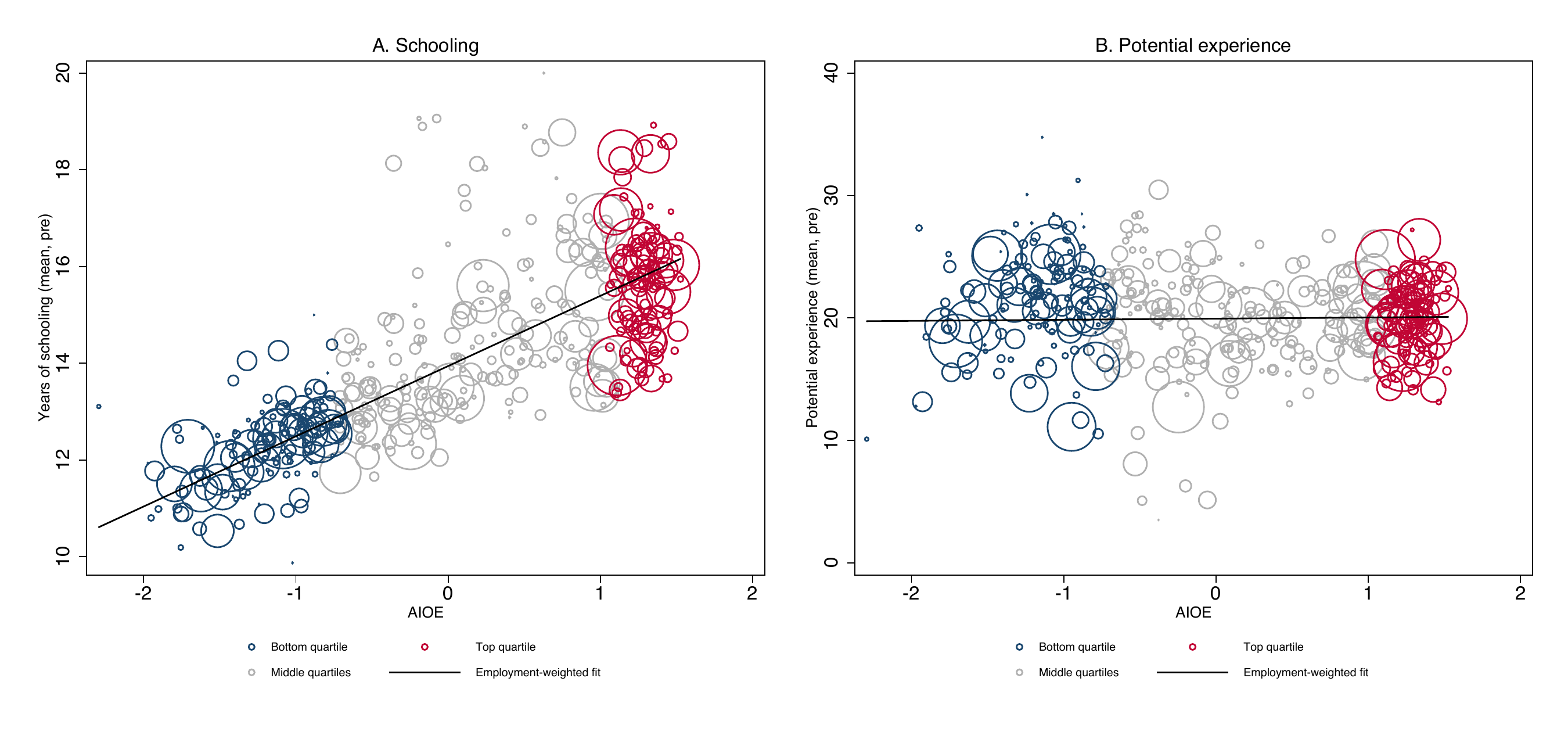}
\caption{Exposure, schooling and experience across occupations}
\label{fig:exposure_schooling}
\begin{minipage}{0.92\textwidth}\footnotesize \textit{Notes:} Each point is an occupational cell; point size is proportional to employment. Averages are computed over the pre-treatment period. The line is an employment-weighted linear fit over all cells.\end{minipage}
\end{figure}
\clearpage

\begin{figure}[H]
\centering
\includegraphics[width=0.88\textwidth]{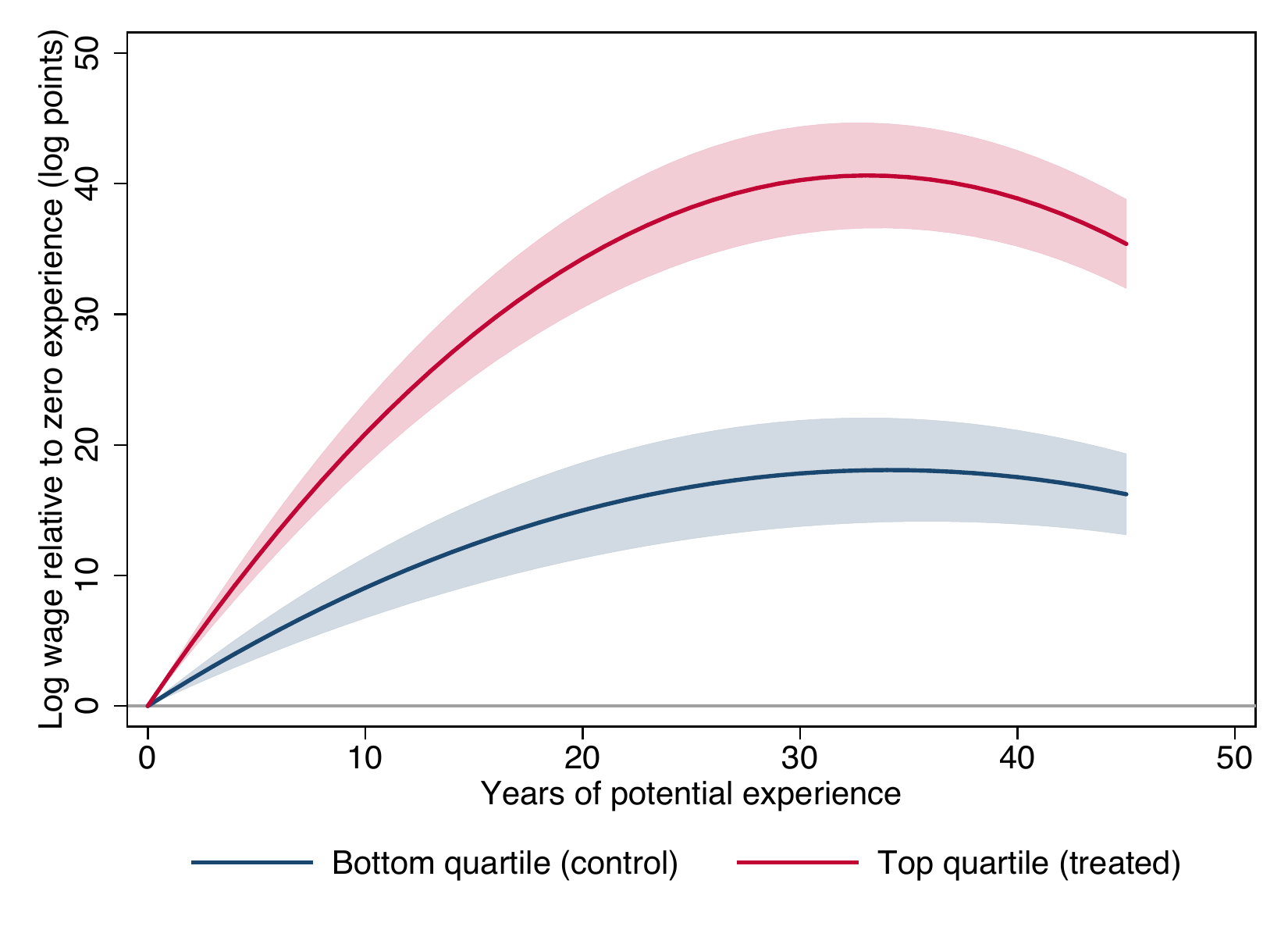}
\caption{Wage-experience profile by exposure group before the shock}
\label{fig:profile_pre}
\begin{minipage}{0.92\textwidth}\footnotesize \textit{Notes:} Predicted log wage from interval regressions estimated separately within each group over the pre-treatment period, normalized to zero at zero years of experience. Shaded areas are 95 percent confidence intervals, with standard errors clustered by occupation.\end{minipage}
\end{figure}
\clearpage

\begin{figure}[H]
\centering
\includegraphics[width=0.88\textwidth]{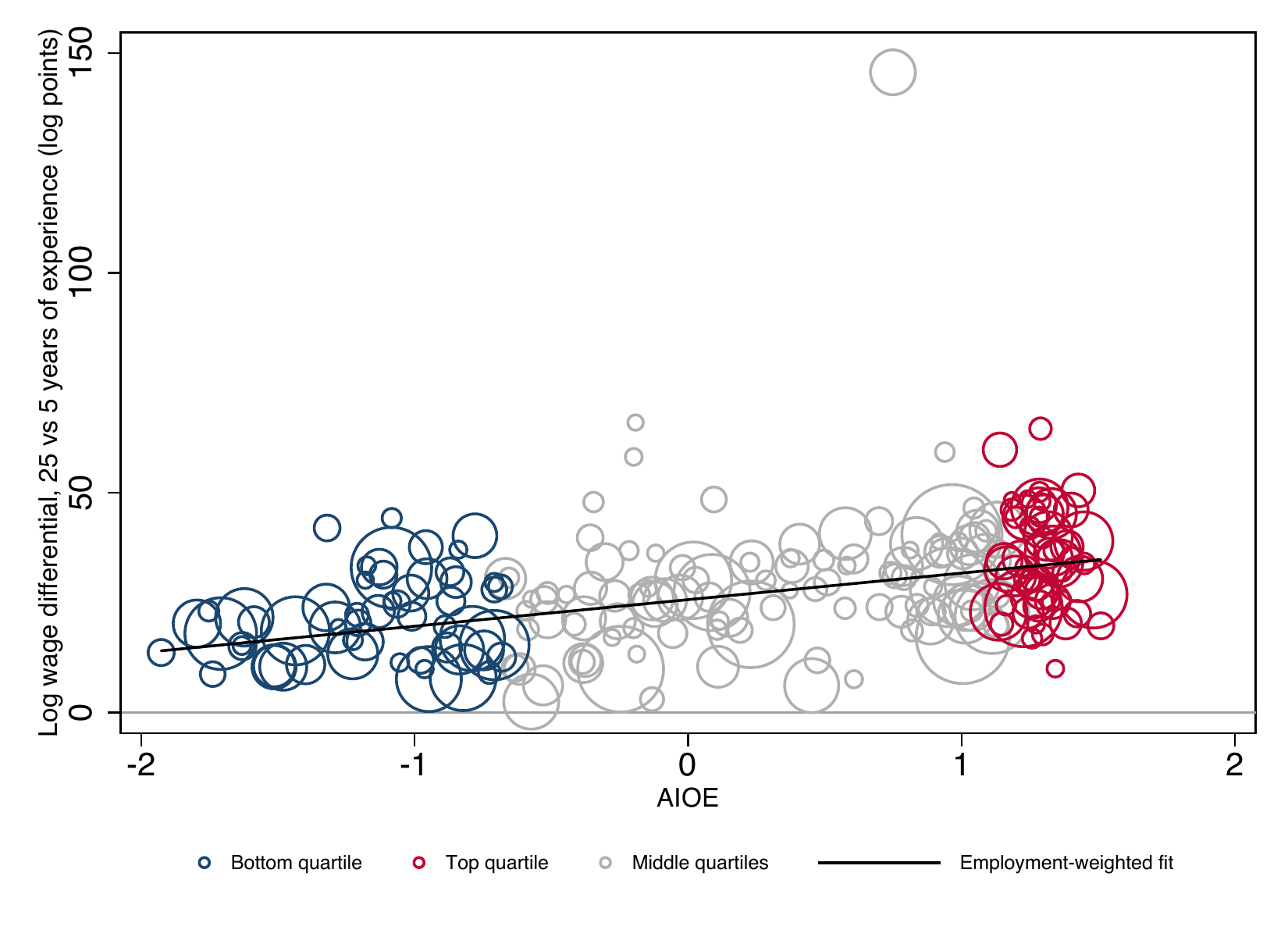}
\caption{The return to experience across occupations}
\label{fig:exp_by_occ}
\begin{minipage}{0.92\textwidth}\footnotesize \textit{Notes:} Each point is an occupational cell with at least 300 observations in the pre-treatment period; point size is proportional to employment. The vertical axis is the log wage differential between 25 and 5 years of potential experience, estimated from a separate interval regression within each cell. The line is an employment-weighted linear fit.\end{minipage}
\end{figure}
\clearpage

\begin{figure}[H]
\centering
\includegraphics[width=0.88\textwidth]{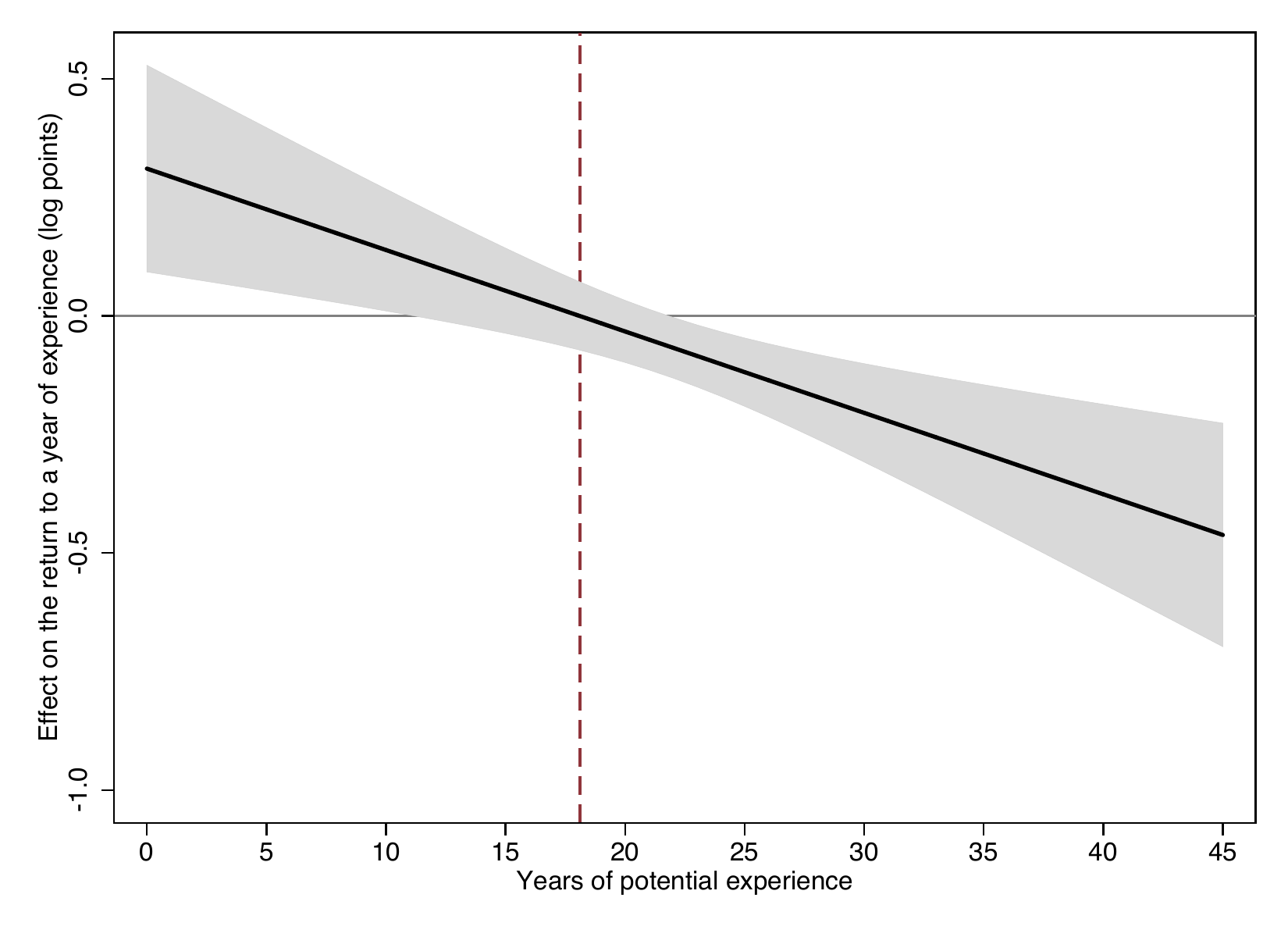}
\caption{The effect on the return to experience across the profile}
\label{fig:profile_effect}
\begin{minipage}{0.92\textwidth}\footnotesize \textit{Notes:} The estimated effect on the return to a year of potential experience, evaluated at every level of experience in the sample. Because the wage equation is quadratic in experience, the effect on the slope of the profile is linear in experience by construction. The shaded area is a 95 percent confidence interval computed from the covariance matrix of the two triple-interaction terms, with standard errors clustered by occupation. The dashed line marks the rotation point, the experience level at which the effect changes sign. Coefficients and confidence intervals are multiplied by 100 and read as log points.\end{minipage}
\end{figure}
\clearpage

\begin{figure}[H]
\centering
\includegraphics[width=0.88\textwidth]{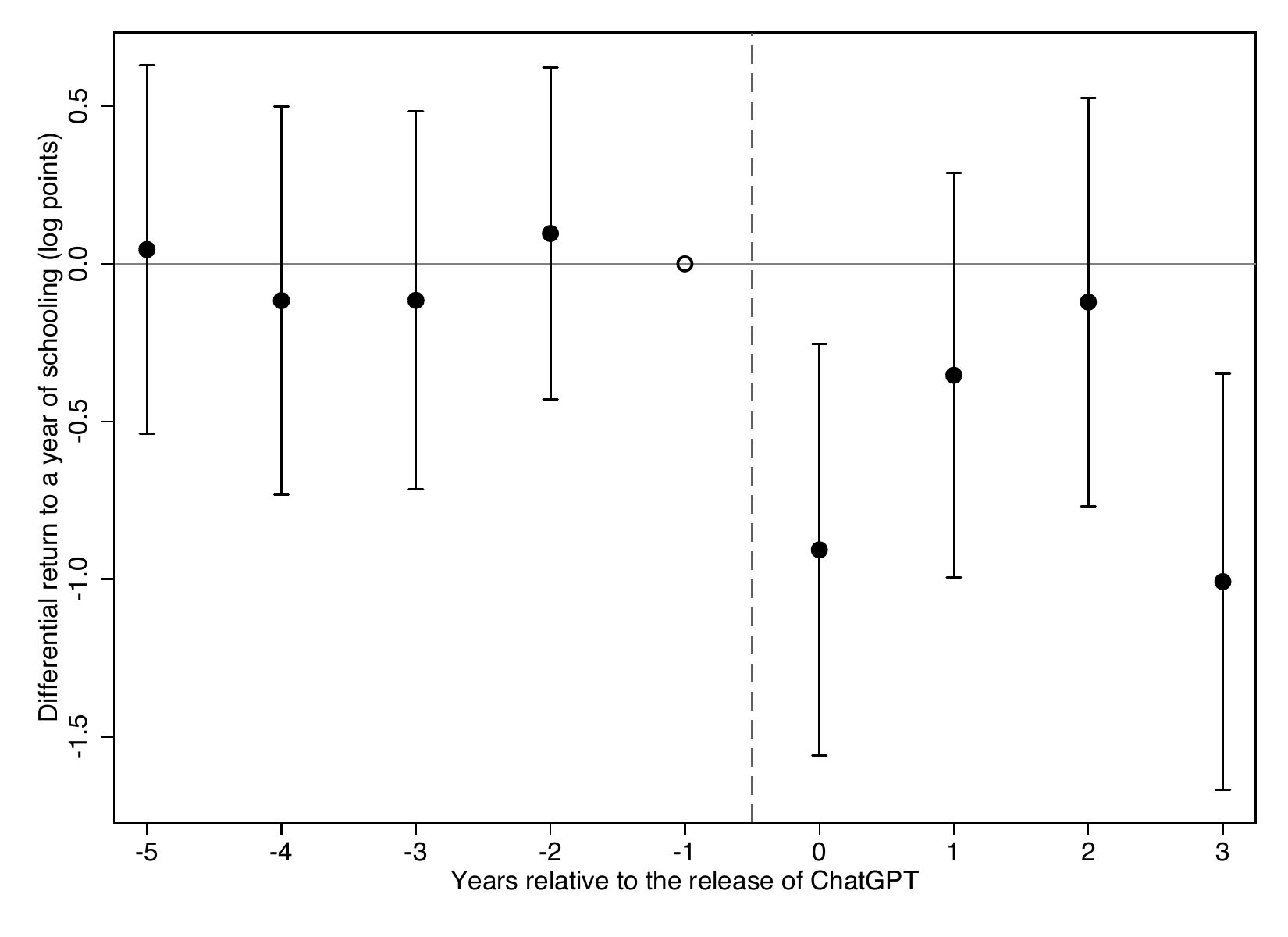}
\caption{The differential return to schooling around the release of ChatGPT}
\label{fig:es_schooling}
\begin{minipage}{0.92\textwidth}\footnotesize \textit{Notes:} Event-study estimates of the differential return to a year of schooling in treated relative to control occupations, by year relative to the release of ChatGPT. Years run from November to October, and the year immediately preceding the release is the omitted category, shown as a hollow marker at zero. Bars are 95 percent confidence intervals with standard errors clustered by occupation. The joint test that the four pre-shock coefficients are zero gives a chi-squared statistic of 1.09 with four degrees of freedom (p = 0.90). Coefficients and confidence intervals are multiplied by 100 and read as log points.\end{minipage}
\end{figure}
\clearpage

\begin{figure}[H]
\centering
\includegraphics[width=0.88\textwidth]{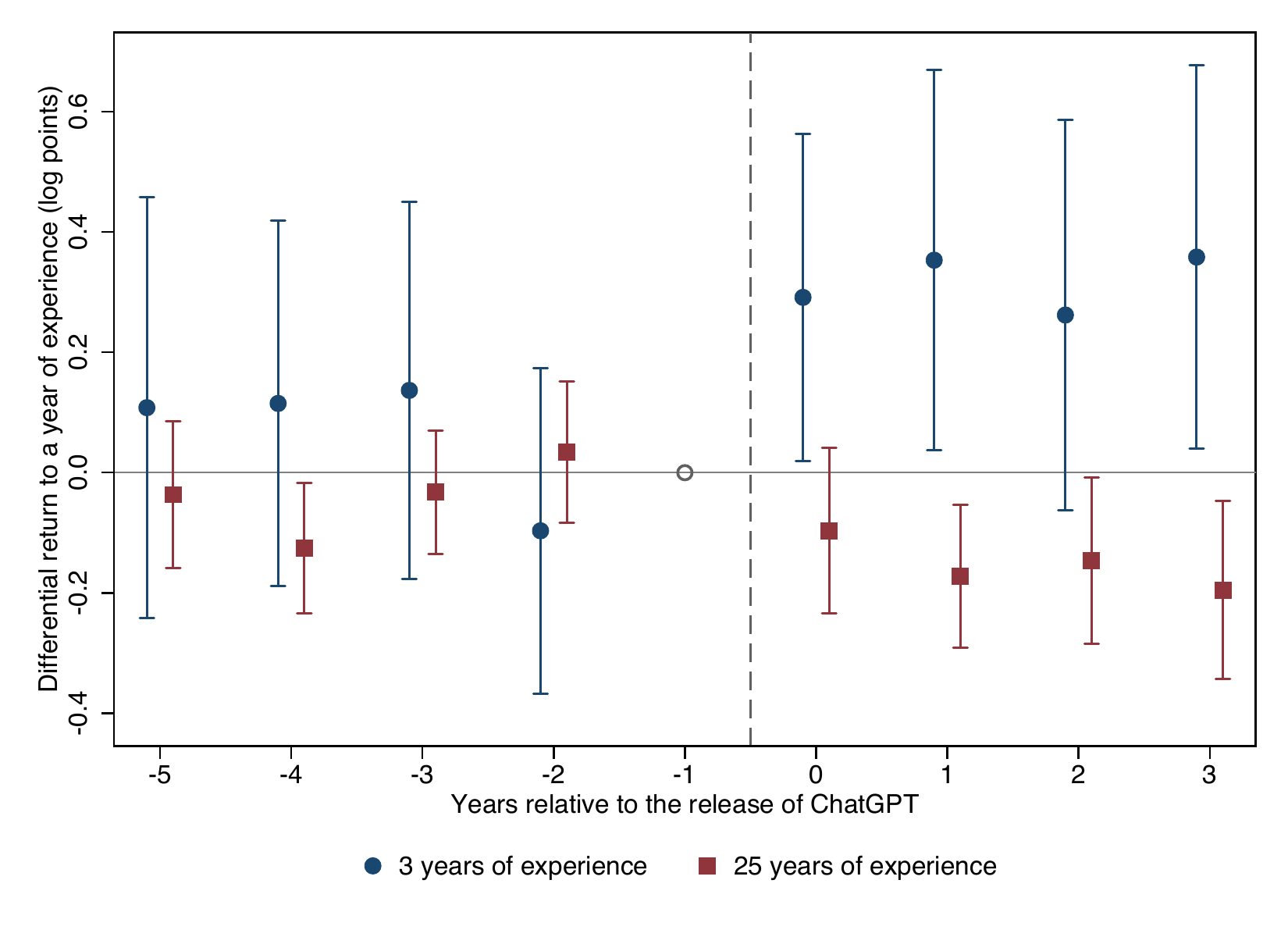}
\caption{The differential return to experience around the release of ChatGPT}
\label{fig:es_experience}
\begin{minipage}{0.92\textwidth}\footnotesize \textit{Notes:} Event-study estimates of the differential return to a year of potential experience in treated relative to control occupations, evaluated at three and at twenty-five years of experience. Each estimate is the derivative of the estimated experience profile at that point. Conventions as in Figure~\ref{fig:es_schooling}. Markers are offset horizontally for legibility. Coefficients and confidence intervals are multiplied by 100 and read as log points.\end{minipage}
\end{figure}
\clearpage

\begin{figure}[H]
\centering
\includegraphics[width=0.88\textwidth]{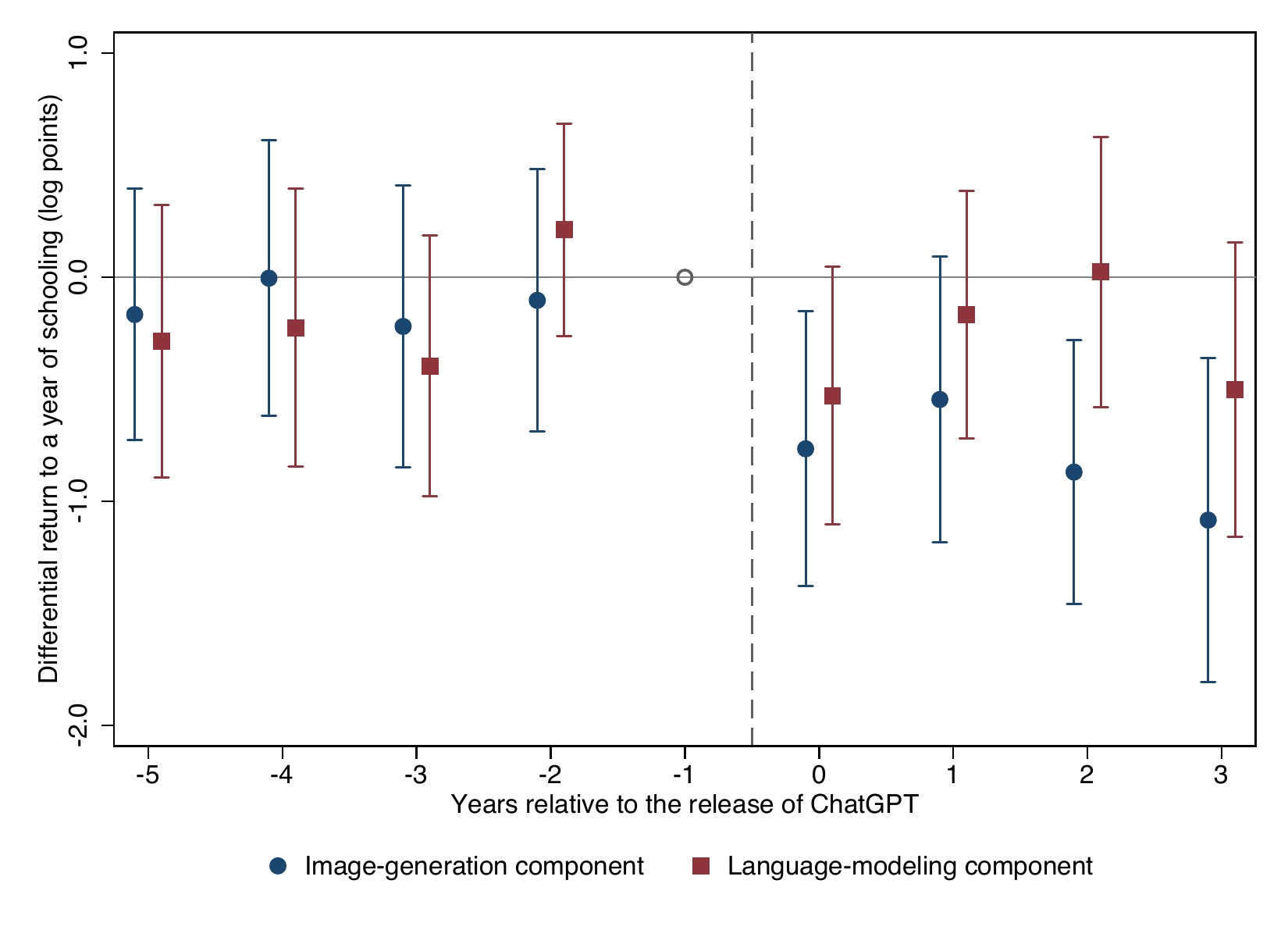}
\caption{The differential return to schooling, by generative component}
\label{fig:es_school_comp}
\begin{minipage}{0.92\textwidth}\footnotesize \textit{Notes:} Event-study estimates of the differential return to a year of schooling when the treatment is defined by quartiles of the image-generation and of the language-modeling components of AIOE, estimated separately. Conventions as in Figure~\ref{fig:es_schooling}. Markers are offset horizontally for legibility. Coefficients and confidence intervals are multiplied by 100 and read as log points.\end{minipage}
\end{figure}
\clearpage

\begin{figure}[H]
\centering
\includegraphics[width=0.88\textwidth]{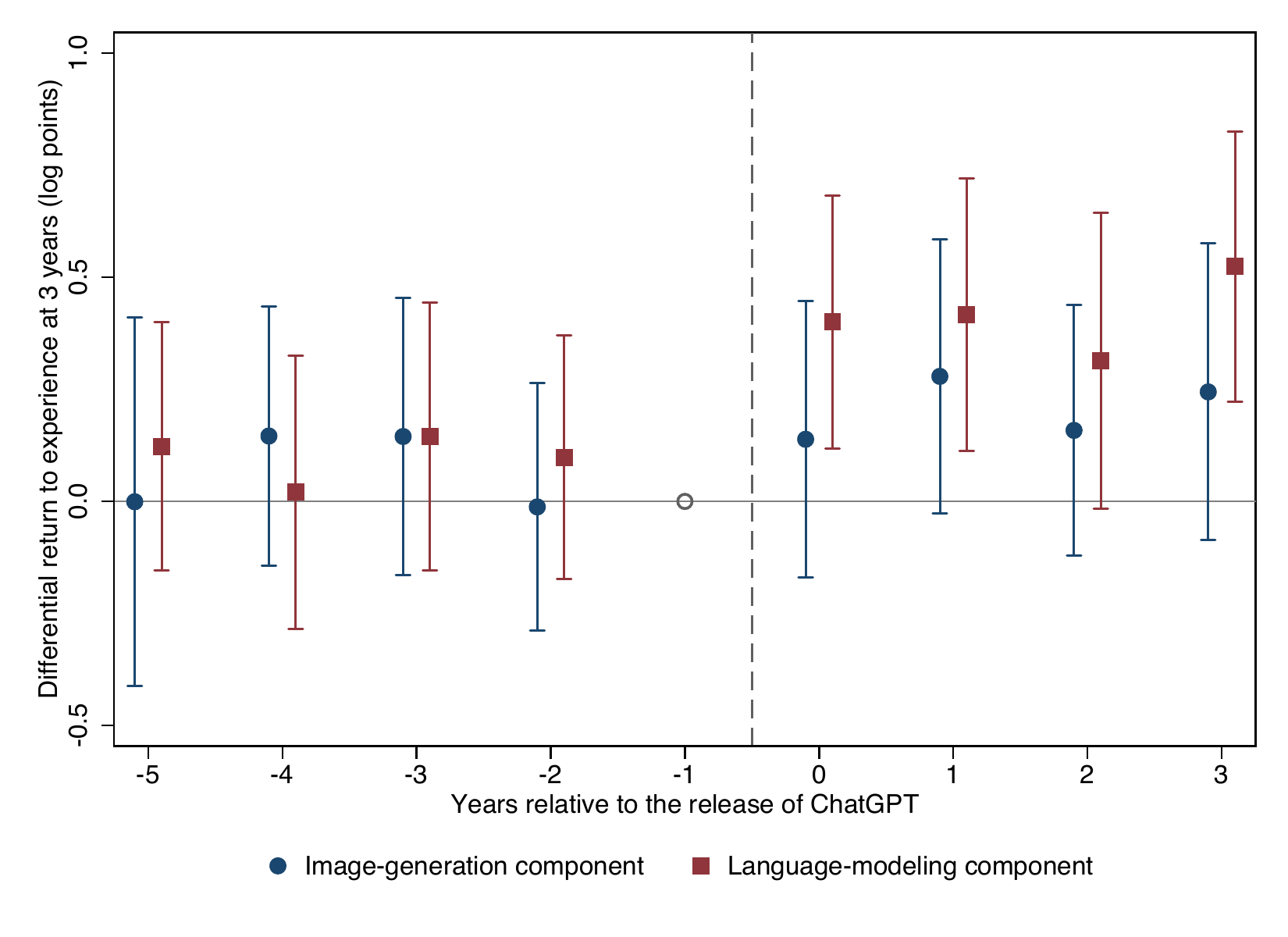}
\caption{The differential return to experience at 3 years, by generative component}
\label{fig:es_exp3_comp}
\begin{minipage}{0.92\textwidth}\footnotesize \textit{Notes:} Event-study estimates of the differential return to a year of potential experience evaluated at three years, by generative component. Conventions as in Figure~\ref{fig:es_experience}. Coefficients and confidence intervals are multiplied by 100 and read as log points.\end{minipage}
\end{figure}
\clearpage
}
\clearpage
\renewcommand{\thetable}{A\arabic{table}}
\setcounter{table}{0}
\renewcommand{\thefigure}{A\arabic{figure}}
\setcounter{figure}{0}

\begin{table}[H]
\centering
\small
\caption{Share of censored observations by exposure group, education and sex}
\label{tab:A_censoring}
\setlength{\tabcolsep}{4pt}
\renewcommand{\arraystretch}{1.12}
\begin{tabular}{@{}>{\raggedright\arraybackslash}p{0.44\textwidth}>{\centering\arraybackslash}p{0.18\textwidth}>{\centering\arraybackslash}p{0.18\textwidth}@{}}
\toprule
 & Control & Treated \\
\midrule
All workers & 0.67 & 11.49 \\
\multicolumn{3}{@{}l}{\textit{Panel A. By education}} \\
Less than high school & 0.24 & 0.90 \\
High school & 0.45 & 2.56 \\
Some college & 0.81 & 3.52 \\
College & 1.72 & 11.47 \\
Postgraduate & 4.19 & 19.13 \\
\multicolumn{3}{@{}l}{\textit{Panel B. By sex}} \\
Men & 0.83 & 17.09 \\
Women & 0.28 & 6.51 \\
\bottomrule
\end{tabular}
\par\medskip
\begin{minipage}{0.98\textwidth}\footnotesize
Notes: Weighted share of observations whose earnings are top-coded, in percent, over the full sample period. Treated (control) occupations are those in the top (bottom) quartile of the AIOE distribution.
\end{minipage}
\end{table}
\clearpage

\begin{table}[H]
\centering
\small
\caption{Quartile cutoffs computed on pre-release data}
\label{tab:A_precutoffs}
\setlength{\tabcolsep}{4pt}
\renewcommand{\arraystretch}{1.12}
\begin{tabular}{@{}>{\raggedright\arraybackslash}p{0.58\textwidth}>{\centering\arraybackslash}p{0.19\textwidth}>{\centering\arraybackslash}p{0.19\textwidth}@{}}
\toprule
 & Baseline & Pre-release cutoffs \\
\midrule
Change in the return to schooling & -0.55*** & -0.56*** \\
 & (0.19) & (0.19) \\
Return to experience, at 3 years & 0.26*** & 0.26*** \\
 & (0.10) & (0.10) \\
at 5 years & 0.23** & 0.22** \\
 & (0.09) & (0.09) \\
at 15 years & 0.05 & 0.05 \\
 & (0.05) & (0.05) \\
at 25 years & -0.12*** & -0.13*** \\
 & (0.04) & (0.04) \\
Experience-profile rotation point (years) & 18.11 & 17.75 \\
 & (2.14) & (2.17) \\
Total effect, 16 years of schooling & -3.15*** & -3.16*** \\
 & (0.81) & (0.81) \\
Pre-trend p, return to schooling & [0.90] & [0.87] \\
Pre-trend p, return to experience at 3 years & [0.55] & [0.64] \\
Pre-trend p, return to experience at 25 years & [0.06] & [0.07] \\
Observations & 315,930 & 318,874 \\
Occupation clusters & 253 & 252 \\
\bottomrule
\end{tabular}
\par\medskip
\begin{minipage}{0.98\textwidth}\footnotesize
Notes: Column 1 reproduces Table~\ref{tab:main}, column 5. Column 2 computes the AIOE quartile cutoffs over January 2018 to October 2022 only and applies them to the full panel. Computer and information systems managers move from the third to the top quartile; pest control workers and coin, vending and amusement machine servicers move from the bottom to the second quartile. Pre-trend rows report, in brackets, p-values of joint tests that the pre-shock event-study coefficients are zero, for the return to schooling and for the return to experience at three and twenty-five years. Standard errors clustered by occupation in parentheses. * p\textless{}0.10, ** p\textless{}0.05, *** p\textless{}0.01. Coefficients and standard errors are multiplied by 100 and read as log points.
\end{minipage}
\end{table}
\clearpage

\begin{table}[H]
\centering
\small
\caption{The ten AI applications underlying the AIOE index and the abilities each one weights}
\label{tab:A_aioe_apps}
\setlength{\tabcolsep}{4pt}
\renewcommand{\arraystretch}{1.12}
\begin{tabular}{@{}>{\raggedright\arraybackslash}p{0.26\textwidth}>{\raggedright\arraybackslash}p{0.70\textwidth}@{}}
\toprule
AI application & Most heavily weighted abilities (weight) \\
\midrule
Abstract strategy games & Information ordering (0.97); Memorization (0.92); Speed of closure (0.90); Flexibility of closure (0.90) \\
Real-time video games & Visual color discrimination (0.94); Information ordering (0.92); Control precision (0.92); Response orientation (0.90) \\
Image recognition & Visual color discrimination (0.93); Near vision (0.91); Spatial orientation (0.89); Far vision (0.89) \\
Visual question answering & Visual color discrimination (0.94); Information ordering (0.93); Near vision (0.91); Spatial orientation (0.89) \\
Generating images & Visualization (0.93); Visual color discrimination (0.90); Information ordering (0.88); Depth perception (0.82) \\
Reading comprehension & Memorization (0.93); Information ordering (0.92); Written comprehension (0.91); Deductive reasoning (0.87) \\
Language modeling & Speech recognition (0.95); Speech clarity (0.93); Oral expression (0.91); Oral comprehension (0.91) \\
Translation & Information ordering (0.89); Oral comprehension (0.89); Written comprehension (0.88); Speech recognition (0.88) \\
Speech recognition & Speech recognition (0.98); Auditory attention (0.91); Oral comprehension (0.89); Speech clarity (0.87) \\
Instrumental track recognition & Auditory attention (0.94); Sound localization (0.89); Hearing sensitivity (0.88); Speech recognition (0.75) \\
\bottomrule
\end{tabular}
\par\medskip
\begin{minipage}{0.98\textwidth}\footnotesize
Notes: Felten, Raj and Seamans build the AIOE index by linking progress in these ten AI applications to the fifty-two O*NET abilities through a crowd-sourced relatedness matrix; the exposure of an occupation is the weighted sum of its ability requirements. For each application, the table lists the four abilities with the largest weights in that matrix. The image-generation and language-modeling components used in Table~\ref{tab:measures} rebuild the index using only the corresponding application.
\end{minipage}
\end{table}
\clearpage

\begin{table}[H]
\centering
\small
\caption{Does the shock change who works in exposed occupations?}
\label{tab:A_composition}
\setlength{\tabcolsep}{4pt}
\renewcommand{\arraystretch}{1.12}
\begin{tabular}{@{}>{\raggedright\arraybackslash}p{0.44\textwidth}>{\centering\arraybackslash}p{0.24\textwidth}@{}}
\toprule
 & Treated $\times$ post \\
\midrule
Age (years) & 0.30** \\
 & (0.14) \\
Years of schooling & 0.02 \\
 & (0.03) \\
Female & 0.53 \\
 & (0.41) \\
Married & 2.53*** \\
 & (0.49) \\
Veteran & 0.27 \\
 & (0.22) \\
Paid hourly & -1.13*** \\
 & (0.39) \\
Hourly with tips or overtime & 0.10 \\
 & (0.32) \\
\bottomrule
\end{tabular}
\par\medskip
\begin{minipage}{0.98\textwidth}\footnotesize
Notes: Each row takes a predetermined worker characteristic as the dependent variable and reports the coefficient on treated $\times$ post, that is, whether the characteristic moved differentially in treated occupations after the shock. The shock cannot change any individual worker's age, schooling or family status, so a non-zero coefficient can only reflect a change in the composition of employment. Age and schooling are in years; the remaining rows are shares, multiplied by 100 and read as percentage points. All regressions include the fixed effects and weights of the wage equation. Standard errors clustered by occupation in parentheses. * p\textless{}0.10, ** p\textless{}0.05, *** p\textless{}0.01.
\end{minipage}
\end{table}
\clearpage

\begin{table}[H]
\centering
\small
\caption{The return to schooling by exposure group and period}
\label{tab:A_returns_group}
\setlength{\tabcolsep}{4pt}
\renewcommand{\arraystretch}{1.12}
\begin{tabular}{@{}>{\raggedright\arraybackslash}p{0.34\textwidth}>{\centering\arraybackslash}p{0.14\textwidth}>{\centering\arraybackslash}p{0.14\textwidth}>{\centering\arraybackslash}p{0.14\textwidth}>{\centering\arraybackslash}p{0.14\textwidth}@{}}
\toprule
 & Pre-shock & Post-shock & Change & Proportional change \\
\midrule
Control occupations & 1.70*** & 1.71*** & 0.01 & 0.6\% \\
 & (0.13) & (0.15) & (0.14) & (8.0) \\
Treated occupations & 7.48*** & 6.94*** & -0.54*** & -7.2\%*** \\
 & (0.45) & (0.38) & (0.15) & (1.7) \\
Difference &  &  & -0.55*** &  \\
 &  &  & (0.19) &  \\
\bottomrule
\end{tabular}
\par\medskip
\begin{minipage}{0.98\textwidth}\footnotesize
Notes: Return to a year of schooling implied by the preferred specification (Table~\ref{tab:main}, column 5), in log points. With the notation of equation (22), the control-group returns are $\beta_S$ before the shock and $\beta_S$ + $\pi_S$ after it; the treated-group returns are $\beta_S$ + $\gamma_S$ and $\beta_S$ + $\gamma_S$ + $\pi_S$ + $\delta_S$. The changes are $\pi_S$ and $\pi_S$ + $\delta_S$, and their difference is $\delta_S$. The proportional change divides each change by the group's pre-shock return; its standard error is computed by the delta method. Standard errors clustered by occupation in parentheses. * p\textless{}0.10, ** p\textless{}0.05, *** p\textless{}0.01.
\end{minipage}
\end{table}
\clearpage

\begin{table}[H]
\centering
\small
\caption{Event-study estimates and joint pre-trend tests}
\label{tab:A_es_full}
\setlength{\tabcolsep}{4pt}
\renewcommand{\arraystretch}{1.12}
\begin{tabular}{@{}>{\raggedright\arraybackslash}p{0.24\textwidth}>{\centering\arraybackslash}p{0.13\textwidth}>{\centering\arraybackslash}p{0.13\textwidth}>{\centering\arraybackslash}p{0.13\textwidth}>{\centering\arraybackslash}p{0.13\textwidth}>{\centering\arraybackslash}p{0.13\textwidth}@{}}
\toprule
 & Return to schooling & \multicolumn{4}{c}{Return to experience, at} \\
\cmidrule(l){3-6}
Year relative to release &  & 3 years & 5 years & 15 years & 25 years \\
\midrule
$-5$ & 0.05 & 0.11 & 0.09 & 0.03 & -0.04 \\
 & (0.30) & (0.18) & (0.16) & (0.08) & (0.06) \\
$-4$ & -0.12 & 0.12 & 0.09 & -0.02 & -0.13** \\
 & (0.31) & (0.16) & (0.14) & (0.07) & (0.06) \\
$-3$ & -0.12 & 0.14 & 0.12 & 0.04 & -0.03 \\
 & (0.31) & (0.16) & (0.14) & (0.07) & (0.05) \\
$-2$ & 0.10 & -0.10 & -0.08 & -0.03 & 0.03 \\
 & (0.27) & (0.14) & (0.12) & (0.07) & (0.06) \\
$-1$ & 0 (omitted) & 0 (omitted) & 0 (omitted) & 0 (omitted) & 0 (omitted) \\
0 & -0.91*** & 0.29** & 0.26** & 0.08 & -0.10 \\
 & (0.33) & (0.14) & (0.12) & (0.06) & (0.07) \\
1 & -0.35 & 0.35** & 0.31** & 0.07 & -0.17*** \\
 & (0.33) & (0.16) & (0.15) & (0.07) & (0.06) \\
2 & -0.12 & 0.26 & 0.22 & 0.04 & -0.15** \\
 & (0.33) & (0.17) & (0.15) & (0.07) & (0.07) \\
3 & -1.01*** & 0.36** & 0.31** & 0.06 & -0.19*** \\
 & (0.34) & (0.16) & (0.15) & (0.08) & (0.08) \\
\midrule
Pre-trend p & [0.90] & [0.55] & [0.59] & [0.84] & [0.06] \\
\bottomrule
\end{tabular}
\par\medskip
\begin{minipage}{0.98\textwidth}\footnotesize
Notes: Event-study estimates of equation~\eqref{eq:eventstudy}, with the year preceding the release of ChatGPT as the omitted category. Years run from November to October. The experience columns report the derivative of the estimated profile at the indicated level of experience. The pre-trend row reports, in brackets, p-values of Wald tests that the four pre-shock coefficients of the corresponding column are zero. Standard errors clustered by occupation in parentheses. * p\textless{}0.10, ** p\textless{}0.05, *** p\textless{}0.01. Coefficients and standard errors are multiplied by 100 and read as log points.
\end{minipage}
\end{table}
\clearpage

\begin{table}[H]
\centering
\small
\caption{Sensitivity of the event-study estimates to violations of parallel trends}
\label{tab:A_honestdid}
\setlength{\tabcolsep}{4pt}
\renewcommand{\arraystretch}{1.12}
\begin{tabular}{@{}>{\raggedright\arraybackslash}p{0.40\textwidth}>{\centering\arraybackslash}p{0.17\textwidth}>{\centering\arraybackslash}p{0.17\textwidth}>{\centering\arraybackslash}p{0.17\textwidth}@{}}
\toprule
 & Estimate, year 0 & Breakdown $\bar{M}$, 95\% & Breakdown $\bar{M}$, 90\% \\
\midrule
Return to schooling & -0.91 & 0.7 & 1.1 \\
Return to experience, at 3 years & 0.29 & 0.1 & 0.3 \\
at 25 years & -0.10 & -- & -- \\
\bottomrule
\end{tabular}
\par\medskip
\begin{minipage}{0.98\textwidth}\footnotesize
Notes: Relative-magnitudes breakdown values from the sensitivity analysis of Rambachan and Roth (2023), for the event-study estimates of Appendix Table~\ref{tab:A_es_full} in the year of the release, the default target of the procedure. Treatment is defined by AIOE as in the baseline. The breakdown value is the largest $\bar{M}$, on a grid of step 0.1, at which the robust confidence set still excludes zero, where $\bar{M}$ bounds each post-shock violation of parallel trends by $\bar{M}$ times the largest violation observed before the shock; 0.0 means the estimate survives only if no violation is allowed, and a dash means the robust confidence set includes zero even then. The two columns report the exercise at the 95 and 90 percent confidence levels. We do not include the credential splits of Section~\ref{sec:heterogeneity-results}: their release-year estimates are not significant at the 5 percent level on any margin, so there is no release-year effect for the exercise to bound. Estimates are multiplied by 100 and read as log points.
\end{minipage}
\end{table}
\clearpage

\begin{table}[H]
\centering
\small
\caption{Event-study estimates with dynamically top-coded wages as point data}
\label{tab:A_pointdata}
\setlength{\tabcolsep}{4pt}
\renewcommand{\arraystretch}{1.12}
\begin{tabular}{@{}>{\raggedright\arraybackslash}p{0.30\textwidth}>{\centering\arraybackslash}p{0.18\textwidth}>{\centering\arraybackslash}p{0.18\textwidth}>{\centering\arraybackslash}p{0.18\textwidth}@{}}
\toprule
 & Return to schooling & \multicolumn{2}{c}{Return to experience, at} \\
\cmidrule(l){3-4}
Year relative to release &  & 3 years & 25 years \\
\midrule
$-5$ & 0.04 & 0.09 & -0.03 \\
  & (0.30) & (0.18) & (0.06) \\
$-4$ & -0.12 & 0.12 & -0.12** \\
  & (0.32) & (0.16) & (0.06) \\
$-3$ & -0.14 & 0.14 & -0.03 \\
  & (0.31) & (0.16) & (0.05) \\
$-2$ & 0.07 & -0.08 & 0.03 \\
  & (0.27) & (0.14) & (0.06) \\
$-1$ & 0 (omitted) & 0 (omitted) & 0 (omitted) \\
0 & -0.74** & 0.38*** & -0.11 \\
  & (0.33) & (0.14) & (0.07) \\
1 & 0.13 & 0.49*** & -0.15*** \\
  & (0.32) & (0.18) & (0.06) \\
2 & 0.48 & 0.39** & -0.11 \\
  & (0.33) & (0.17) & (0.07) \\
3 & -0.38 & 0.57*** & -0.18** \\
  & (0.42) & (0.17) & (0.07) \\
\midrule
Pre-trend p & [0.91] & [0.59] & [0.07] \\
\bottomrule
\end{tabular}
\par\medskip
\begin{minipage}{0.98\textwidth}\footnotesize
Notes: Event-study estimates of equation~\eqref{eq:eventstudy} in which wages top-coded under the dynamic regime enter as point observations at their Census-assigned values, the mean of the top three percent of earners, rather than as censored intervals; top-coded wages of the static era remain intervals. This is the event-study counterpart of Table~\ref{tab:estimation}, column 4. Because the Census assigns a single value per month, this variant removes the variation within the upper tail that the interval likelihood exploits, so it serves as a stress test of the tail model rather than as an alternative estimate. Conventions as in Appendix Table~\ref{tab:A_es_full}. Standard errors clustered by occupation in parentheses. * p\textless{}0.10, ** p\textless{}0.05, *** p\textless{}0.01. Coefficients and standard errors are multiplied by 100 and read as log points.
\end{minipage}
\end{table}
\clearpage

\begin{table}[H]
\centering
\small
\caption{Functional form of the experience profile}
\label{tab:A_funcform}
\setlength{\tabcolsep}{4pt}
\renewcommand{\arraystretch}{1.12}
\begin{tabular}{@{}>{\raggedright\arraybackslash}p{0.34\textwidth}>{\centering\arraybackslash}p{0.145\textwidth}>{\centering\arraybackslash}p{0.145\textwidth}>{\centering\arraybackslash}p{0.145\textwidth}>{\centering\arraybackslash}p{0.145\textwidth}@{}}
\toprule
 & Quadratic (baseline) & Restricted cubic spline, 5 knots & Restricted cubic spline, 4 knots & Quartic \\
\midrule
Return to experience, at 3 years & 0.26*** & 0.33 & 0.21 & 0.37 \\
 & (0.10) & (0.22) & (0.16) & (0.27) \\
at 5 years & 0.23** & 0.30 & 0.20 & 0.28 \\
 & (0.09) & (0.19) & (0.15) & (0.18) \\
at 15 years & 0.05 & -0.03 & 0.08 & 0.01 \\
 & (0.05) & (0.14) & (0.06) & (0.09) \\
at 25 years & -0.12*** & -0.03 & -0.11* & -0.09 \\
 & (0.04) & (0.13) & (0.07) & (0.08) \\
at 30 years & -0.20*** & -0.16* & -0.21*** & -0.14 \\
 & (0.05) & (0.08) & (0.06) & (0.10) \\
at 35 years & -0.29*** & -0.31*** & -0.27*** & -0.22** \\
 & (0.07) & (0.10) & (0.09) & (0.09) \\
Experience at which the effect changes sign (years) & 18.1 & 13.4 & 19.3 & 15.9 \\
Change in the return to schooling & -0.55*** & -0.54*** & -0.53*** & -0.54*** \\
 & (0.19) & (0.20) & (0.20) & (0.19) \\
\bottomrule
\end{tabular}
\par\medskip
\begin{minipage}{0.98\textwidth}\footnotesize
Notes: Each column re-estimates the preferred specification of Table~\ref{tab:main}, column 5, replacing the quadratic in potential experience with the indicated function, entered on its own and interacted with treatment, post, and treatment $\times$ post. The restricted cubic splines place knots at the percentiles of potential experience recommended by Harrell (2001): the 5th, 27.5th, 50th, 72.5th and 95th with five knots (1, 10, 19, 30 and 42 years) and the 5th, 35th, 65th and 95th with four (1, 13, 26 and 42 years). Each row reports the derivative of the treatment $\times$ post component of the profile at the indicated experience. Standard errors clustered by occupation in parentheses. * p\textless{}0.10, ** p\textless{}0.05, *** p\textless{}0.01. Coefficients and standard errors are multiplied by 100 and read as log points.
\end{minipage}
\end{table}
\clearpage

\begin{table}[H]
\centering
\small
\caption{The experience profile within credential groups}
\label{tab:A_credential}
\setlength{\tabcolsep}{4pt}
\renewcommand{\arraystretch}{1.12}
\begin{tabular}{@{}>{\raggedright\arraybackslash}p{0.44\textwidth}>{\centering\arraybackslash}p{0.20\textwidth}>{\centering\arraybackslash}p{0.20\textwidth}@{}}
\toprule
 & Without college credential & With college credential \\
\midrule
Change in the return to schooling & -0.81 & -0.91* \\
 & (0.86) & (0.50) \\
Return to experience, at 3 years & 0.30** & 0.43*** \\
 & (0.13) & (0.16) \\
at 5 years & 0.27** & 0.38*** \\
 & (0.12) & (0.15) \\
at 15 years & 0.11* & 0.13* \\
 & (0.07) & (0.07) \\
at 25 years & -0.04 & -0.12* \\
 & (0.05) & (0.07) \\
Observations & 160,064 & 155,866 \\
Occupation clusters & 248 & 250 \\
\bottomrule
\end{tabular}
\par\medskip
\begin{minipage}{0.98\textwidth}\footnotesize
Notes: The preferred specification of Table~\ref{tab:main}, column 5, estimated separately for workers without a college credential and for workers with an associate degree or higher. Because potential experience is age minus schooling minus six, workers with the same potential experience but different credentials differ in age; estimating within credential groups compares workers at similar points of comparable careers. The rotation of the experience profile is present within both groups and is strongest among credentialed workers. Standard errors clustered by occupation in parentheses. * p\textless{}0.10, ** p\textless{}0.05, *** p\textless{}0.01. Coefficients and standard errors are multiplied by 100 and read as log points.
\end{minipage}
\end{table}
\clearpage

\begin{figure}[H]
\centering
\includegraphics[width=0.98\textwidth]{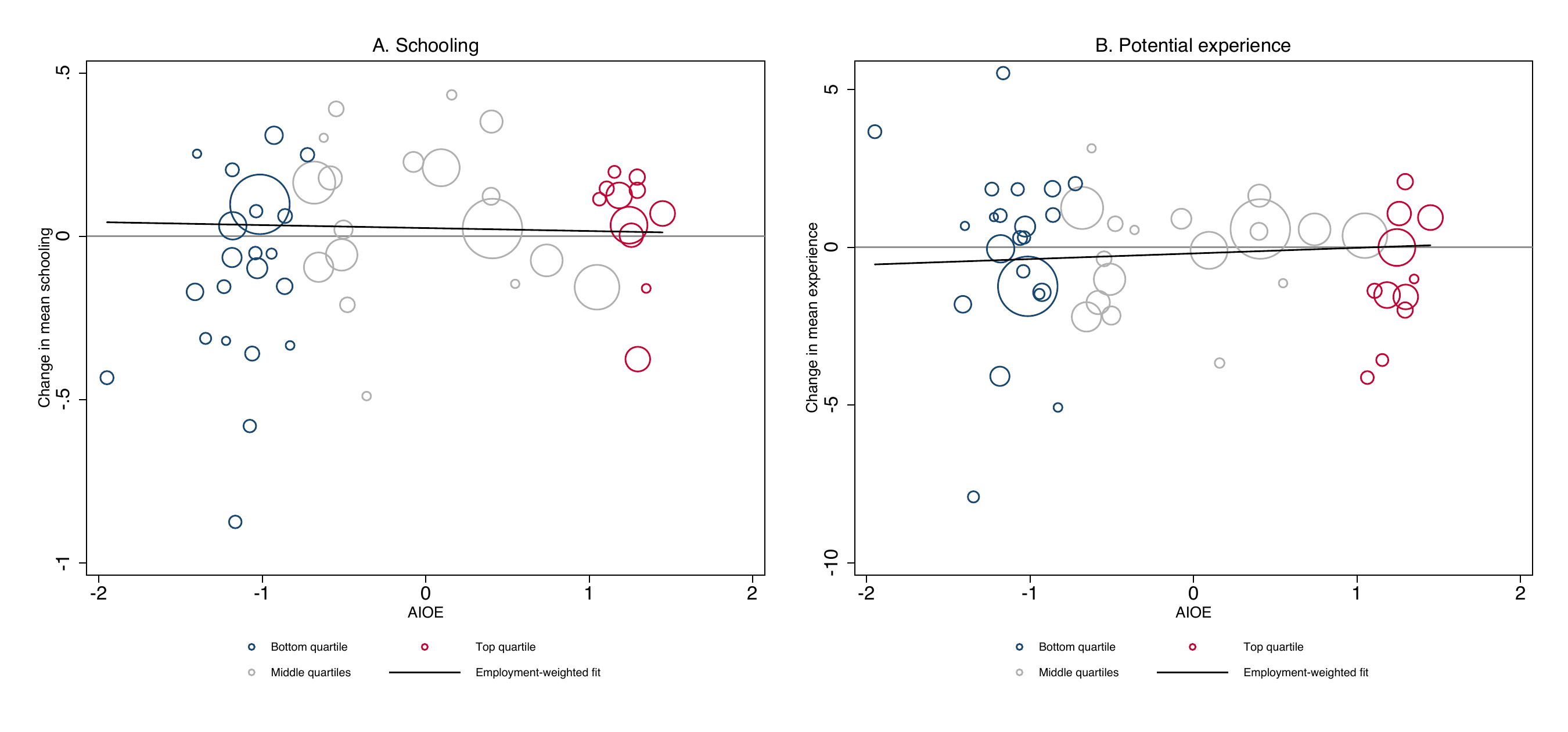}
\caption{Changes in occupational composition and exposure}
\label{fig:A_composition}
\begin{minipage}{0.92\textwidth}\footnotesize \textit{Notes:} Each point is an occupational cell with employment above the tenth percentile in both periods; point size is proportional to pre-treatment employment. The vertical axis is the change in the cell's average schooling (panel A) and average potential experience (panel B) between the pre- and post-treatment periods. Employment-weighted slopes on exposure are $-0.009$ (s.e. 0.027) and 0.178 (s.e. 0.225), respectively.\end{minipage}
\end{figure}
\clearpage

\begin{figure}[H]
\centering
\includegraphics[width=0.88\textwidth]{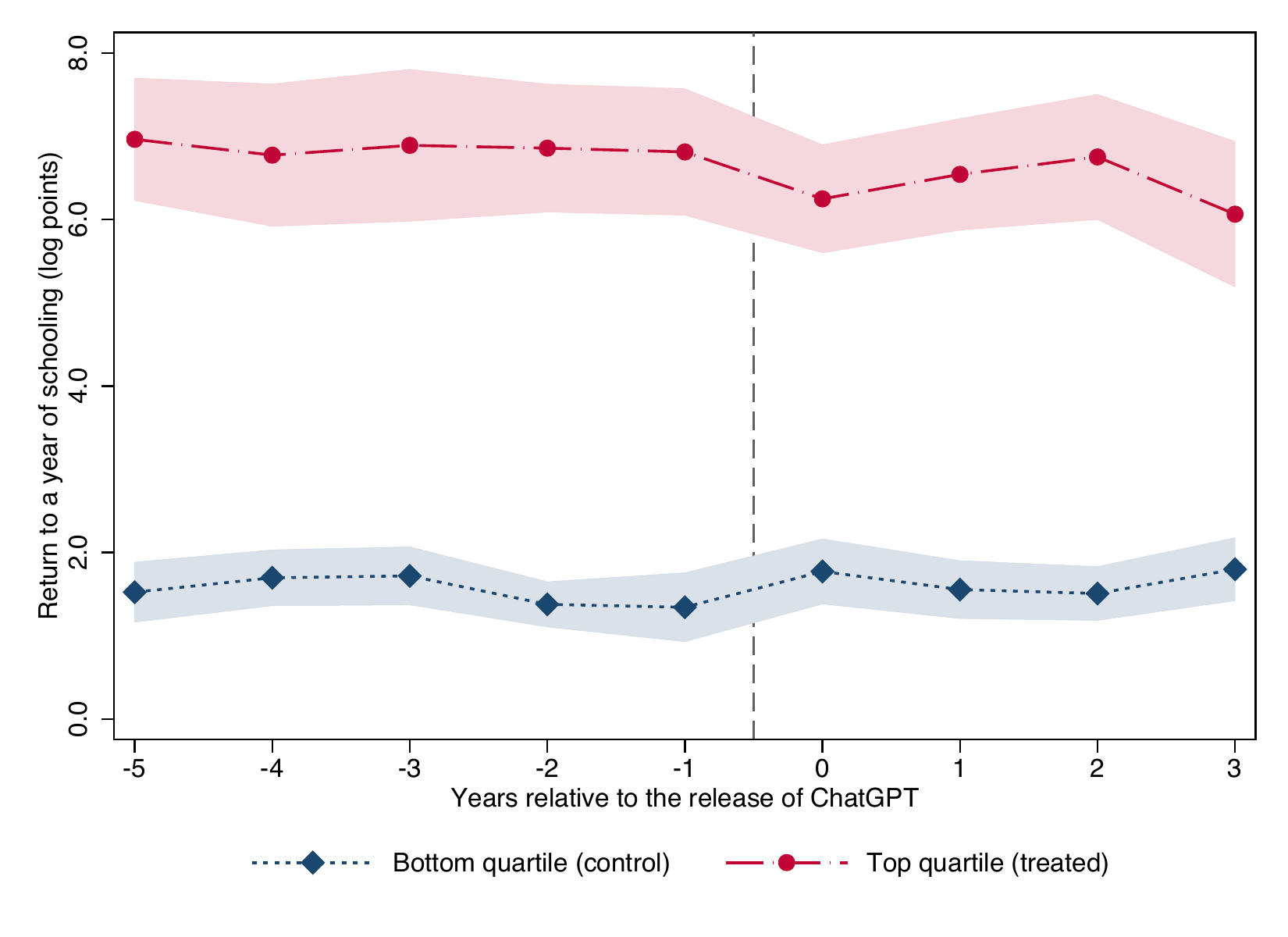}
\caption{The return to schooling by exposure group over time}
\label{fig:A_returns_group_time}
\begin{minipage}{0.92\textwidth}\footnotesize \textit{Notes:} Coefficient on years of schooling from interval regressions estimated separately within each group and each year relative to the release of ChatGPT, with occupation, state and month fixed effects. Years run from November to October; the vertical line marks November 2022. Shaded areas are 95 percent confidence intervals, with standard errors clustered by occupation. Because the regressions include occupation fixed effects, the coefficients measure the return to schooling within occupations. Coefficients and confidence intervals are multiplied by 100 and read as log points.\end{minipage}
\end{figure}
\clearpage

\begin{figure}[H]
\centering
\includegraphics[width=0.88\textwidth]{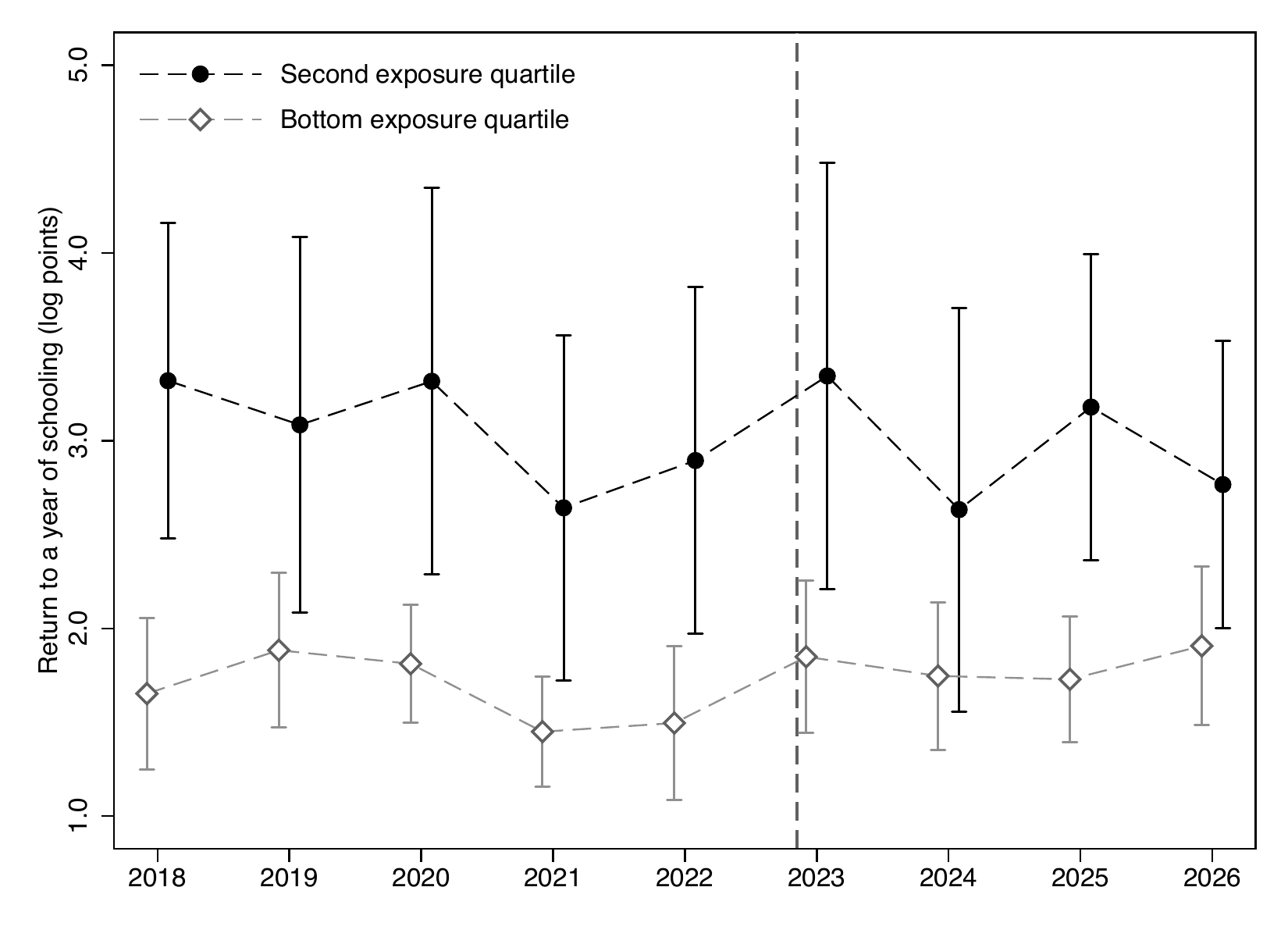}
\caption{The return to schooling in the two untreated quartiles over time}
\label{fig:A_q2q1}
\begin{minipage}{0.92\textwidth}\footnotesize \textit{Notes:} Coefficient on years of schooling by quartile from one interval regression per calendar year, allowing the returns to schooling and experience to differ across quartiles, with occupation, state and month fixed effects and the controls of the baseline specification. The sample contains the second and bottom exposure quartiles only, neither of which is treated. Bars are 95 percent confidence intervals, with standard errors clustered by occupation. The vertical line marks November 2022. Coefficients and confidence intervals are multiplied by 100 and read as log points.\end{minipage}
\end{figure}
\clearpage

\begin{figure}[H]
\centering
\includegraphics[width=0.88\textwidth]{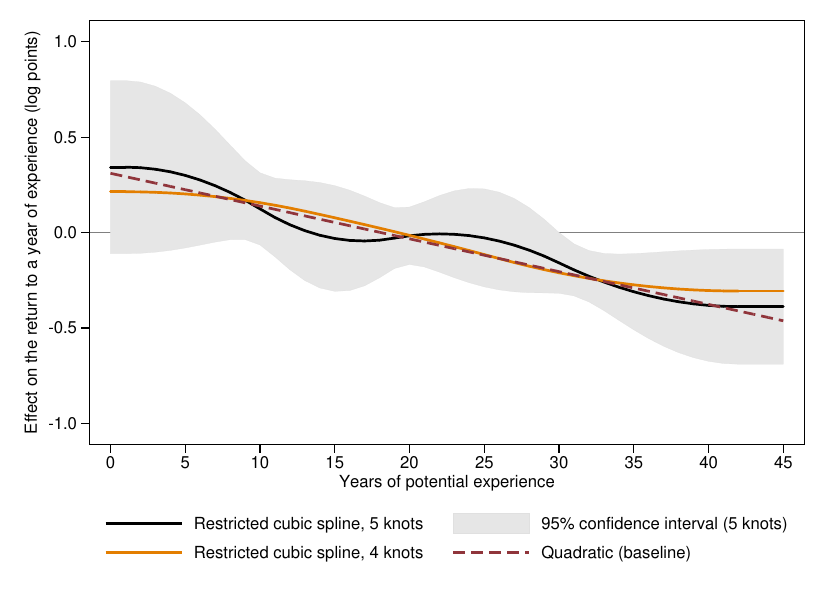}
\caption{The effect on the return to experience under flexible functional forms}
\label{fig:A_funcform}
\begin{minipage}{0.92\textwidth}\footnotesize \textit{Notes:} Effect of the release on the return to a year of potential experience in treated relative to control occupations, evaluated at every level of experience. The solid black line uses a restricted cubic spline with five knots at Harrell's (2001) percentiles, and the shaded area is its 95 percent confidence interval with standard errors clustered by occupation. The orange line uses four knots, and the dashed line is the baseline quadratic of Figure~\ref{fig:profile_effect}. Coefficients and confidence intervals are multiplied by 100 and read as log points.\end{minipage}
\end{figure}
\clearpage
\end{document}